\documentclass[aip, jcp, reprint]{revtex4-2}

\usepackage{amsfonts,amsmath,amssymb,bm}
\usepackage{graphicx,color,url,orcidlink}
\usepackage[english]{babel}

\begin{document}

\title{Non-markovian neural quantum propagator and its application to the simulation of ultrafast nonlinear spectra}

%%%%%%%%%%
%% Remember to add the orcid
%%%%%%%%%%
\author{Jiaji Zhang\,\orcidlink{0000-0003-2978-274X}}
\email{jiaji.zhang@zhejianglab.com} 
\affiliation{Zhejiang Laboratory, Hangzhou 311100, China}

\author{Lipeng Chen\,\orcidlink{0009-0002-1541-8912}}
\email{chenlp@zhejianglab.com}
\affiliation{Zhejiang Laboratory, Hangzhou 311100, China}

%%%%%%%%%%%%
% abstract
\begin{abstract}
The accurate solution of dissipative quantum dynamics plays an important role on the simulation of open quantum systems. Here we propose a  machine-learning-based universal solver for the hierarchical equations of motion,
one of the most widely used approaches which takes into account non-markovian effects and nonperturbative system-environment interactions in a numerically exact manner. We develop a neural quantum propagator model by utilizing the neural network architecture, which avoids time-consuming iterations and can be used to evolve any initial quantum state for arbitrarily long times. To demonstrate the efficacy of our model, we apply it to the simulation of population dynamics and linear and two-dimensional spectra of the Fenna-Matthews-Olson complex.
\end{abstract}

\maketitle

%%%%%%%%%%%%
% introduction
\section{Introduction}
\label{sec.intro}

% Ultrafast nonlinear spectroscopy

Ultrafast nonlinear spectroscopy provides a versatile tool to reveal the electronic structure and chemical reaction mechanism. \cite{gelin2022chemrev, nisoli2017chemrev, mukamel2000arpc, maiuri2019jacs, dorfman2016rmv}
Two-dimensional electronic spectroscopy (2DES), in particular, has been widely employed to monitor electronic excitation dynamics in polyatomic molecules. {\cite{fresch2023nrmp, oliver2018rsos, cohen2011chemphys, ginsberg2009acr}} 
By utilizing multiple UV-Vis pulses, one can measure the correlation between different electronic states via off-diagonal peaks of 2DES. 
In addition, extracting dynamical information from the evolution of 2DES enables the direct visualization of chemical reaction processes. {\cite{scholes2011natchem, martin2011jacs, arsenault2021jcp, kim2020natcomm, ruetzel2013prl}}

% Theoretical simulation of spectroscopy
Theoretical simulation of nonlinear spectra is based on the response function theory, which requires the accurate description of system dynamics upon interaction with external laser pulses.{\cite{cho2019, mukamel1999}} 
As molecular systems inevitably interact with their surrounding environment, a commonly used strategy is to treat the environmental degrees of freedom as a heat bath, and derive the equations of motion for the reduced system after tracing out bath degrees of freedom. {\cite{breuer2007, weiss2012}} 
The hierarchical equations of motion (HEOM) is one of the best-known quantum dynamics approaches, which takes into account non-markovian effects and non-perturbative system-environment interactions in a numerically exact manner. {\cite{tanimura2020jcp, ye2016wirescms, zhang2022imheom}} 
As a typical partial differential equation (PDE), one recursively solves HEOM by conventional iterative solvers such as fourth-order Runge-Kutta (RK4) and split-operator methods. {\cite{kloeden1992, yan2021jcp, ke2023jcp}} 
Despite their straightforward implementation, the main drawbacks of iterative methods are the large computational cost and long-time numerical instabilities. 
While improvements have been proposed to alleviate the numerical issues of iterative methods, an efficient universal solver is still to be proposed. {\cite{kimura2014, schlimgen2021prl, liu2023jcp}}

% Machine learning, surrogate models 
Over recent years, the fast development of machine learning technique offers new possibilities to   circumvent aforementioned difficulties. {\cite{lecun2015nature, hermann2023natrevchem}}
A variety of machine learning based surrogate models have been developed to provide universal solvers for PDE. {\cite{lu2021don, li2021fno, kovachki2023fno, guibas2021afno}} 
In contrast to iterative methods, those surrogate models solve the PDE by defining a functional that describes the mapping between an arbitrary initial condition and its corresponding solution at some subsequent time. 
This functional is then parameterized as a deep neural network and optimized with a prepared dataset. 
The state-of-the-art surrogate models, such as Fourier Neural Operator (FNO)  and DeepONet, have shown their effectiveness over conventional methods on a set of PDEs of the classical dynamical problems.{\cite{lu2021deepxde, jaideep2022fcn, jiang2021dte}}

% Contribution of this work
In this work, we extend the surrogate models to the non-markovian quantum dynamics by developing a so-called neural quantum propagator (NQP) model for HEOM. 
As the quantum analogue of the universal PDE solver, the NQP model directly generates dynamics of system without invoking the tedious, expensive iterations. 
Following our previous work, we adopt the FNO architecture as the core neural network structure. {\cite{zhang2024jpcl}} 
We test its performance by comparing with the conventional RK4 method in various computational scenarios. 
In addition to the simulation of population dynamics, we also employ the NQP model to compute linear and nonlinear spectra. 

% Super resolution algorithm
Similar to other neural network architectures, training NQP model requires a large amount of high precision data. 
Those data can only be generated by conventional iterative solvers with a small enough time step, which in turn leads to a large computational cost in the data preparation stage. 
To address this issue, we introduce a super-resolution algorithm, which only relies on the low-resolution data to construct high-resolution NQP. 
The intrinsic error in the dataset is systematically improved by utilizing the physics-informed loss function (PILF), which is defined directly from the HEOM. 
The optimization of PILF does not rely on any prepared dataset, which significantly improves the overall computational performance of the NQP model.

% Structure of this paper
The rest of the paper is organized as follows. In Section \ref{sec.theory}, we introduce the HEOM approach and linear and nonlinear response functions.
In Section \ref{sec.neural}, we present the NQP model, including the FNO architecture, the training setup, and the super-resolution algorithm.
Numerical demonstrations on the Fenna-Matthews-Olson (FMO) system are presented in Section \ref{sec.result}. 
Finally, conclusions are drawn in Section \ref{sec.conclusion}.

%%%%%%%%%%%%
% dissipative theory
\section{Methodology}
\label{sec.theory}

% HEOM, FMO system
\subsection{HEOM approach}

% Hamiltonian
We consider an electronic system interacting with a set of heat baths.
The total Hamiltonian can be written as
\begin{equation}
\hat{H}_{tot} = \hat{H}_{s} + \hat{H}_{b} + \hat{H}_{s-b}.
\label{eq.H_total}
\end{equation}
Here, the first term $\hat{H}_{s}$ is the Hamiltonian of the electronic system,
\begin{equation}
\hat{H}_{s} = \sum_{j=1}^{N} \varepsilon_{j} | j\rangle \langle j | + 
\sum_{j \ne j^{\prime}}\Delta_{j, j^{\prime}} | j\rangle \langle j^{\prime} | ,
\end{equation}
where $\varepsilon_{j}$ is the energy of the $j$-th electronic state $|j\rangle$, 
and $\Delta_{j, j^{\prime}}$ is the interstate coupling.
The second term is the Hamiltonian of harmonic heat baths,
\begin{equation}
\hat{H}_{b} = \sum_{j=1}^{N} \sum_{\nu} \left( \frac{\hat{p}_{j,\nu}^{2}}{2} 
+ \frac{\omega_{j,\nu}^{2} \hat{x}_{j,\nu}^{2}}{2}\right),
\end{equation}
where $\hat{p}_{j,\nu}$, $\hat{x}_{j, \nu}$, and $\omega_{j, \nu}$ 
are the dimensionless momentum, coordinate, and frequency of $\nu$-th oscillator of $j$-th bath.
The last term is the system-bath interaction Hamiltonian,
\begin{equation}
\hat{H}_{s-b} = - \sum_{j=1}^{N} \hat{V}_{j} \sum_{\nu} g_{j,\nu} \hat{x}_{j,\nu},
\end{equation}
where $\hat{V}_{j}=|j\rangle\langle{j}|$, and
$g_{j,\nu}$ is the coupling constant between the $j$-th state and the $\nu$-th oscillator which can be specified by a spectral density,
\begin{equation}
J_j(\omega)=\sum_{\nu}g_{j,\nu}^2\delta(\omega-\omega_{j,\nu}).
\end{equation}

% HEOM equation under high-temperature limit
The influence of the $j$-th heat bath on the electronic system is characterized by the bath correlation function,
{\cite{breuer2007, weiss2012}}
\begin{eqnarray}
\begin{aligned}
&C_{j}(t) \\
= &\frac{1}{\pi} \int_{0}^{\infty} {\rm{d}}\omega J_{j}(\omega) 
\left[ \coth\left(\frac{\beta\hbar\omega}{2}\right) \cos(\omega t) - i \, \sin(\omega t) \right],
\end{aligned}
\label{eq.bath_corr_func}
\end{eqnarray}
where $J_{j}(\omega)$ is the spectral density of the $j$-th bath,
$\beta = 1/k_{B} T$ is the inverse temperature with $k_{B}$ being the Boltzmann constant. We model the bath by the Drude spectral density,
\begin{equation}
J_{j}(\omega) = \frac{2\lambda_{j} \gamma_{j} \omega}{\gamma_{j}^{2}  + \omega^{2}},
\end{equation}
where $\lambda_{j}$ is the reorganization energy, and $\gamma_{j}$ is the inverse of the bath correlation time.
In this paper, we consider the high-temperature approximation ($\beta\hbar\gamma_{j} <1$), and express Eq. {\eqref{eq.bath_corr_func}} as $C_{j}(t) = c_{j} e^{-\gamma_{j} |t|} $, where 
\begin{equation}
c_{j} = \frac{2 \lambda_{j}}{\beta\hbar^{2}} - i \, \frac{\lambda_{j}\gamma_{j}}{\hbar}.
\end{equation}
To go beyond this approximation, one can include so-called low-temperature correction terms. {\cite{ishizaki2005jpsj, hu2010jcp}} The time evolution of the reduced density matrix can be described by the HEOM approach, which is written as
{\cite{tanimura2006jpsj, tanimura2020jcp}}
\begin{equation}
\begin{aligned}
\partial_t  \hat{\rho}_{\vec{n}}(t) &= - \left[ \frac{i}{\hbar} \hat{H}_{s}^{\times} + \sum_{j=1}^{N} n_{j} \gamma_{j} \right] \hat{\rho}_{\vec{n}}(t) 
- i\sum_{j=1}^{N} \hat{V}_{j}^{\times} \hat{\rho}_{\vec{n}+\vec{e}_{j}}(t) \\
&- i \sum_{j=1}^{N} \left[ c_{j} \hat{V}_{j}  \hat{\rho}_{\vec{n}-\vec{e}_{j}}(t) -
c_{j}^{\ast}  \hat{\rho}_{\vec{n}-\vec{e}_{j}}(t) \hat{V}_{j}  \right],
\end{aligned}
\label{eq.heom}
\end{equation}
where $\vec{n} = \{ n_1, n_2, ..., n_{N} \}$ denotes the index vector with $n_{j}$ being the non-negative integer, and we have introduced abbreviated notations,
$\hat{A}^{\times} \hat{B} = \hat{A}\hat{B} - \hat{B} \hat{A}$. The density operator with all indexes equal to zero, $\hat{\rho}_{\vec{0}}(t)$ with $\vec{0} = \{ 0, 0, ..., 0 \}$,  corresponds to
the density operator of the reduced electronic system, 
while all other density operators are introduced to describe non-markovian and non-perturbative effects.

% spectroscopy and response functions
\subsection{Linear and nonlinear response functions}

The linear and nonlinear spectra are evaluated within the framework of response function theory.
{\cite{mukamel1999, gelin2022chemrev, zhang2023jcp}}
The linear response function is defined as
\begin{equation}
R^{(1)}(t) = \left( \frac{i}{\hbar} \right) {\rm{Tr}} \left\{ \hat{\mu} 
\mathcal{G}_{tot}(t) \hat{\mu}^{\times} \hat{\rho}_{tot}(0) \right\},
\label{eq.resp_linear}
\end{equation}
where $\hat{\mu}$ is the transition dipole operator, and $\hat{\rho}_{tot}$ and
$\mathcal{G}_{tot}(t) = \exp( - i \hat{H}_{tot}^{\times}/\hbar t)$ are 
the density operator and the quantum propagator of the total system, respectively.
The linear absorption spectrum is obtained by the Fourier transformation
\begin{equation}
R^{(1)}(\omega) = {\rm{Im}} \int_{0}^{\infty} {\rm{d}}t e^{i \omega t} R^{(1)}(t) ,
\label{eq.resp_linear_fft}
\end{equation} 
where Im denotes the imaginary part.
The third-order response function is defined as
\begin{equation}
\begin{aligned}
R^{(3)}&(t_3, t_2, t_1) = \left( \frac{i}{\hbar} \right)^{3}\\
&{\rm{Tr}} \left\{ \hat{\mu} \mathcal{G}_{tot}(t_3) \hat{\mu}^{\times}
\mathcal{G}_{tot}(t_2) \hat{\mu}^{\times} \mathcal{G}_{tot}(t_1) 
\hat{\mu}^{\times} \hat{\rho}_{tot}(0) \right\}.
\end{aligned}
\label{eq.resp_3rd}
\end{equation}
The rephasing and non-rephasing parts of 2D spectrum are defined by
\begin{equation}
\begin{aligned}
R^{(3, R)}&(\omega_3, \omega_1; t_2) = {\rm{Im}} \\
&\int_{0}^{\infty} {\rm{d}}t_3 \int_{0}^{\infty} {\rm{d}}t_1 
e^{i \omega_3 t_3 - i \omega_1 t_1} R^{(3)}(t_3, t_2, t_1) ,
\end{aligned}
\label{eq.resp_3rd_reph}
\end{equation}
\begin{equation}
\begin{aligned}
R^{(3, NR)}&(\omega_3, \omega_1; t_2) = {\rm{Im}} \\
&\int_{0}^{\infty} {\rm{d}}t_3
\int_{0}^{\infty} {\rm{d}}t_1 e^{i \omega_3 t_3 + i \omega_1 t_1} R^{(3)}(t_3, t_2, t_1) ,
\end{aligned}
\label{eq.resp_3rd_nonreph}
\end{equation}

% evaluation
Within the HEOM formalism, Eqs. {\eqref{eq.resp_linear}} and {\eqref{eq.resp_3rd}} 
can be evaluated by replacing $\hat{\rho}_{tot}$ and $\mathcal{G}_{tot}(t)$ with  $\hat{\rho}_{\vec{n}}(0)$ and Eq. {\eqref{eq.heom}}, respectively. The final trace is only taken for the zeroth order element of $\hat{\rho}_{\vec{n}}(t)$, i.e., $\hat{\rho}_{\vec{0}}(t)$.

%%%%%%%%%%%%
% neural network model
\section{Neural quantum propagator}
\label{sec.neural}

% architecture
We introduce the abbreviated index, $x = (j, j^{\prime}, n_1, n_2, ..., n_{N})$, and align the matrix entries $\rho(x, t) = \langle j | \hat{\rho}_{\vec{n}}(t) | j^{\prime} \rangle$ as the column vector
\begin{equation}
\vec{\rho}_{t} = \{ \rho(x_0, t), \rho(x_1, t), ...  \}.
\label{eq.heom_adv}
\end{equation}
The HEOM (Eq. {\eqref{eq.heom}}) can be recast to a matrix-vector form as 
\begin{equation}
\partial_t \vec{\rho}_{t} = {\bm{L}} \,  \vec{\rho}_{t},
\label{eq.heom_matvec}
\end{equation}
where the matrix entries of ${\bm{L}}$ can be inferred from the right-hand side of Eq. {\eqref{eq.heom}}.
The propagator of HEOM is then defined through the integration form as ${\bm{G}}_{t} = \exp({t {\bm{L}}})$, 
which satisfies the composition property,
\begin{equation}
 \vec{\rho}_{t} = {\bm{G}}_{t - t_0} \vec{\rho}_{t_0} = 
 e^{(t - t_0) {\bm{L}}}.
\label{eq.comp_prop}
\end{equation}
To facilitate the description of later sections, we also introduce uniform time grid as $t_m = m \delta_t$ for $m = 1 \sim N_t$, where $\delta_t = t_{max} / N_t$ is the time step with $N_t$ and $t_{max}$ being the total number of time steps and the fixed upper time limit, respectively.

% FNO
\subsection{Model's architecture}

\begin{figure}
\centering
\includegraphics[width=0.5\textwidth]{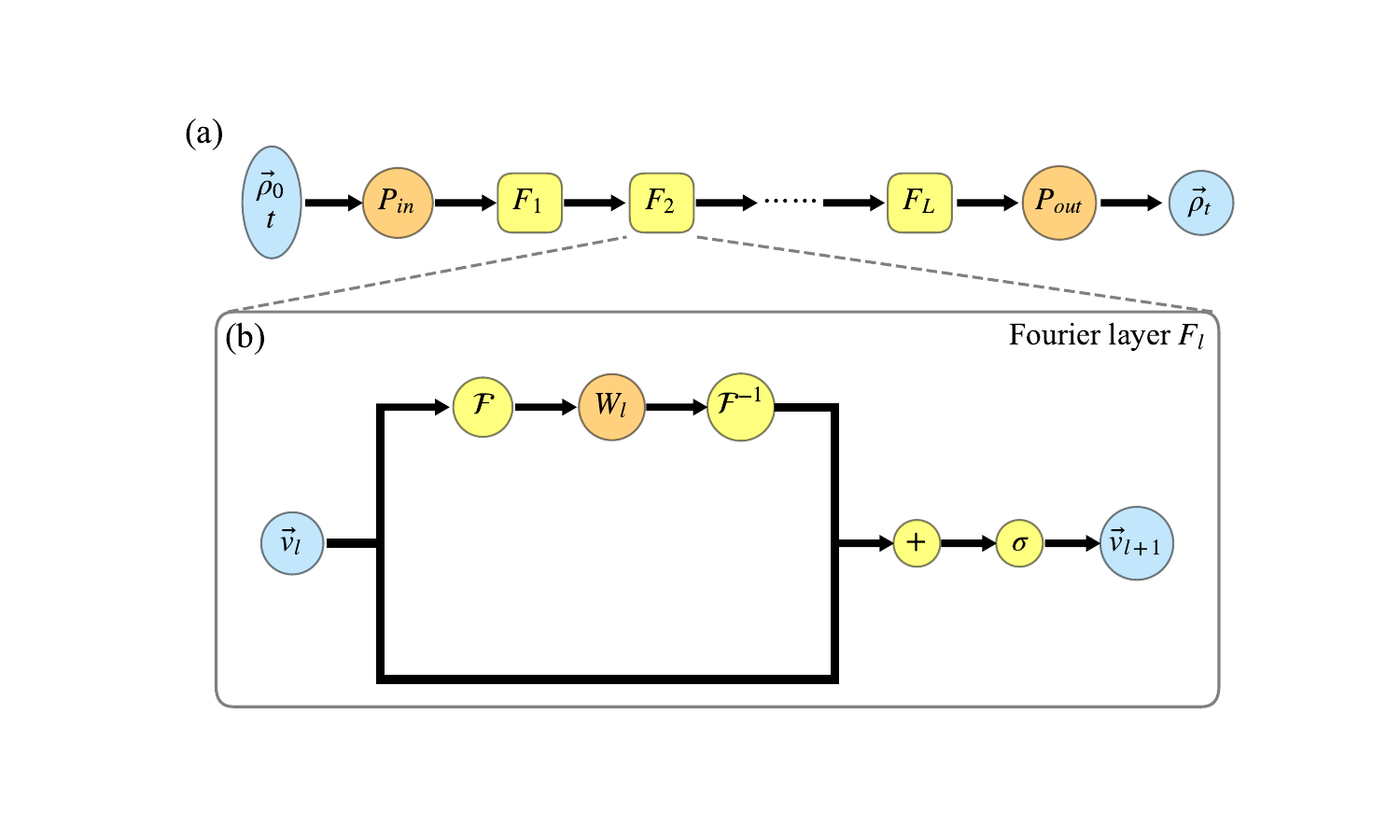}
\caption{The architecture of (a) the NQP model, and (b) the $l$-th Fourier layer.
Here, $\mathcal{F}$ and $\mathcal{F}^{-1}$ denote the Fourier transform and its inverse. 
$+$ and $\sigma$ represent the element-wise sum and the GeLU activation function. 
The learnable parameters are those in $P_{in}$, $P_{out}$, and $W_{l}$.}
\label{fig.model}
\end{figure}

The NQP model serves as an extension of FNO to the quantum dynamics. 
Its mathematical foundation is the extended universal approximation theorem, 
which states that large enough neural network can approximate the functional,
representing the mapping between input and output pairs. {\cite{kovachki2021fno}} To construct the NQP model, we follow our previous work {\cite{zhang2024jpcl}}
and parameterize the HEOM propagator as a deep neural network, ${\bm{G}}_{t_m}[\theta]$, where $\theta$ represents all the trainable parameters.
The architecture of the NQP model is shown in Fig. {\ref{fig.model}}.
In Fig. {\ref{fig.model}}(a), $P_{in}$ and $P_{out}$ are the linear projections between physical and latent Fourier spaces.
They are parameterized as the point-wise convolution network with one hidden layer 
and a Gaussian Error Linear Unit (GeLU) activation function.
The rest parts are called the Fourier layers with their structure presented in Fig. {\ref{fig.model}}(b).

To process the input of the $l$-th layer $\vec{v}_{l}$, two different routes are adopted.
On the upper route, ${\mathcal{F}}$ and ${\mathcal{F}}^{-1}$ denote the Fourier and its inverse transform.  
The point-wise convolution $W_{l}$ serves as the learnable weight in Fourier space.
Only the lowest $k_{max}$ modes are explicitly included in the weight tensor, 
while others with higher frequencies are truncated to control the size of the model and avoid the numerical instabilities. 
The lower route is similar to the residual network. 
The results of two different routes are summed and activated by GeLU before passing to the next layer.

% Model architecture

The NQP model takes all the entries of initial condition $\vec{\rho}_{0}$ and a chosen time $t$ as the input, and outputs $\vec{\rho}_{t} = {\bm{G}}_{t}[\theta]\vec{\rho}_{0}$ satisfying Eq. {\eqref{eq.heom_matvec}}.
It should be noted that no restrictions are a priori made on the explicit forms of $\vec{\rho}_{0}$. The NQP model can be directly applied to the simulation of response function by taking the field interaction form $\hat{\mu}^{\times}\hat{\rho}_{\vec{n}}$ as the input.
Since the composition property is also retained during the parameterization, the time evolution up to arbitrarily long times can be obtained by recursively applying Eq. {\eqref{eq.comp_prop}}.

% loss function
\subsection{Training objective}

The NQP model is trained by minimizing an objective function $\mathcal{L}$, defined as
\begin{equation}
\mathcal{L} = \alpha \mathcal{L}_{data} + (1 - \alpha)\mathcal{L}_{phys},
\label{eq.loss_def}
\end{equation}
where $\mathcal{L}_{data}$ and $\mathcal{L}_{phys}$ are referred to as the data and physics-informed loss functions, respectively. 
The hyper-parameter $\alpha \in (0, 1)$ serves as a weight factor, which will be dynamically adjusted in the training stage.

% Data loss
For the data part $\mathcal{L}_{data}$, we prepare a dataset by randomly sampling a set of initial condition $\{ \vec{\rho}_{0} \}$, 
and then evaluating their time evolution $\{\vec{\rho}_{t} \}$ up to $t \in [0, t_{max}]$ using conventional RK4 method.
The data loss function is defined as follows,
\begin{equation}
\mathcal{L}_{data} = \sum_{p=1}^{N_{data}} \sum_{m=1}^{N_t} 
\frac{{\left| \left|  {\bm{G}}_{t_m}[\theta]\vec{\rho}_{0}^{(p)} 
- \vec{\rho}_{t_m}^{(p)} \right|\right|}_{F}}
{{\left| \left| \vec{\rho}_{t_m}^{(p)} \right|\right|}_{F}},
\end{equation}
where $||\cdot||_{F}$ denotes the Frobenius-norm, $N_{data}$ is the number of individual samples in the dataset, and
$\vec{\rho}_{0}^{(p)}$ and $\vec{\rho}_{t_m}^{(p)}$ are the initial condition and the
corresponding evolution for the $p$-th sample, respectively.

% physics-informed loss
To ensure the universality of ${\bm{G}}_{t}[\theta]$ that is applicable to any $\vec{\rho}_{0}$, one needs a large number of samples $N_{data}$, which leads to even more computational cost in the data preparation stage. We introduce a physics-informed loss function to reduce the effective number of samples $N_{data}$ while keeping the universality of ${\bm{G}}_{t}[\theta]$.{\cite{rosofsky2023mlst}} 
The physics-informed loss function is defined by minimizing the 
difference between left- and right-hand sides of Eq. {\eqref{eq.heom_matvec}} as
\begin{equation}
\mathcal{L}_{phys} = \sum_{p^{\prime}=1}^{N_{phys}}\sum_{m=1}^{N_t} 
{\left| \left| {\partial_t} {\bm{G}}_{t_m}[\theta]\vec{\rho}_{0}^{(p^{\prime})} 
- {\bm{L}} {\bm{G}}_{t_m}[\theta] \vec{\rho}_{0}^{(p^{\prime})} \right|\right|}_{F},
\end{equation}
where $N_{phys}$ is the number of samples in the physics dataset. 
The time derivative $\partial_t \vec{\rho}_{t}$ is evaluated by the finite difference method. It should be mentioned that the calculation of 
$\mathcal{L}_{phys}$ involves far less samples  as compared to that of $\mathcal{L}_{data}$. 
In addition, we adopt the on-the-fly sampling approach by re-generating the physics dataset 
at each training epoch to further improve the performance of the trained model.

% super resolution
\subsection{Super resolution algorithm}

To further reduce the computational cost in the data preparation stage, we introduce a super resolution algorithm, which allows the construction of the high resolution NQP model from a lower resolution dataset. 
As illustrated from the previous subsection, the lower resolution dataset is prepared by integrating the HEOM with a larger time step $K \delta_t$ ($K > 1$) for a set of $\{ \vec{\rho}_{0} \}$ using the RK4 method. 
The obtained data is then embedded into the finer grid $\{ t_{m} = m \delta_t \}$ by interpolating the missing value using the linear interpolation scheme. $\mathcal{L}_{data}$ is evaluated on this interpolated dataset in the training stage.

On the other hand, the physics-informed loss function $\mathcal{L}_{phys}$ is evaluated directly on the finer time grid and serves as the correction over the deviation from the dataset. 
The super resolution algorithm is then completed by dynamically adjusting the weight factor $\alpha$ in Eq. {\eqref{eq.loss_def}} during the training process.
At the begining, we set $\alpha = 1$ and randomly initilize all the model's parameters. 
During the training process, $\alpha$ is gradually decreased to a small enough value such as $\sim 0.01$, and $\mathcal{L}_{phys}$ gradually becomes the dominant contribution term.
The minimization of $\mathcal{L}_{phys}$ allows the improvement of the resolution over the intrinsic deviation of the dataset.

% Why data is necessary
At the end of this subsection, we briefly discuss the possibility of data-free training, which is achieved by fixing $\alpha = 0$ and using only $\mathcal{L}_{phys}$ during the training process. 
From a theoretical point of view, training with or without $\mathcal{L}_{data}$ results in the same model as long as $\mathcal{L}_{phys}$ becomes the dominant contribution of Eq. {\eqref{eq.loss_def}}.
In practice, however, training with only $\mathcal{L}_{phys}$ requires  longer epochs for convergence when all the learnable parameters are randomly initialized. 
In this case, a prepared dataset, even with low resolution, serves as a well-performed guidance for the training.

%%%%%%%%%%%%
% result
\section{Numerical experiments}
\label{sec.result}

In the following, we use seven-sites FMO-complex (apo-FMO) 
as model system and adopt the Adolph and Renger's Hamiltonian. {\cite{adolphs2006bj, ishizaki2009pnas}} We restrict the discussion to the one-exciton manifold. The electronic state $|j \rangle$ ($j=1,\cdots,7$) corresponds to the state where only $j$-th pigment is excited, 
and $| g\rangle $ denotes the ground state without any excitation. 
The system-bath interaction is chosen as $\hat{V}_{j} = |j\rangle \langle j|$ and $\hat{V}_{g} \equiv 0$.
The heat bath parameters are chosen as $\lambda_{j} = 35 \, {\rm{cm}}^{-1}$, $\gamma_{j} = 200 \, {\rm{cm}}^{-1}$, and $T = 300 \, {\rm{K}}$, respectively. The HEOM is truncated at the hierarchy level of $\sum n_{j} \le 2$ after adopting the filtering algorithm.{\cite{shi2009jcp}} 
Within our choice of parameters and under the high-temperature limit, we found that it is accurate enough for the testing of our NQP model.
We set the upper time limit as $t_{max} = 30 \, {\rm{fs}}$ with a time step of $\delta_t = 0.6 \, {\rm{fs}}$, which results in $N_{t} = 50$ time points.

% Numerical details: system parameters, training details 
\subsection{Training and validation test}

We first introduce the model's hyper-parameters and training setup.
In order to train the NQP model, we prepare the low-resolution training dataset by randomly sampling $N_{data} = 3000$ initial conditions $\vec{\rho}_{0}$. 
The low resolution dataset is prepared by integrating HEOM with a larger time step of $3 \delta_t$.
The missing values are linearly interpolated when embedded into the finer grid with the time step of $\delta_t$.
To test the accuracy of the model, we also prepare a high-resolution validation set with 500 samples, following the same setup but using a smaller time step of $\delta_t$. It should be noted that the high-resolution validation set is never referred in the training stage.
In the training process, the physics dataset is prepared using the on-the-fly sampling algorithm by randomly generating $N_{phys} = 2000$ initial conditions at each epoch.

The other hyper-parameters of the NQP model are chosen as follows.
We set the hidden channel of projections $P_{in}$ and $P_{out}$ as $512$.
We use $4$ Fourier layers, each of which has a hidden channel of size $64$, and the total number of trainable parameters is around 10 million.
The model is trained for $10^{5}$ epochs using the Adam optimizer.
The learning rate is initially set to $10^{-4}$, and then halved every 500 epochs until reaching $~10^{-6}$.
The weight factor $\alpha$ in Eq. {\eqref{eq.loss_def}} is initialized as $\alpha = 1$, and halved every 100 epochs until reaching $\sim 10^{-2}$.
All the tasks are performed on the Nvidia A40 GPU with 48 GB memory.

% Model architecture
\begin{figure}
\centering
\includegraphics[width=0.5\textwidth]{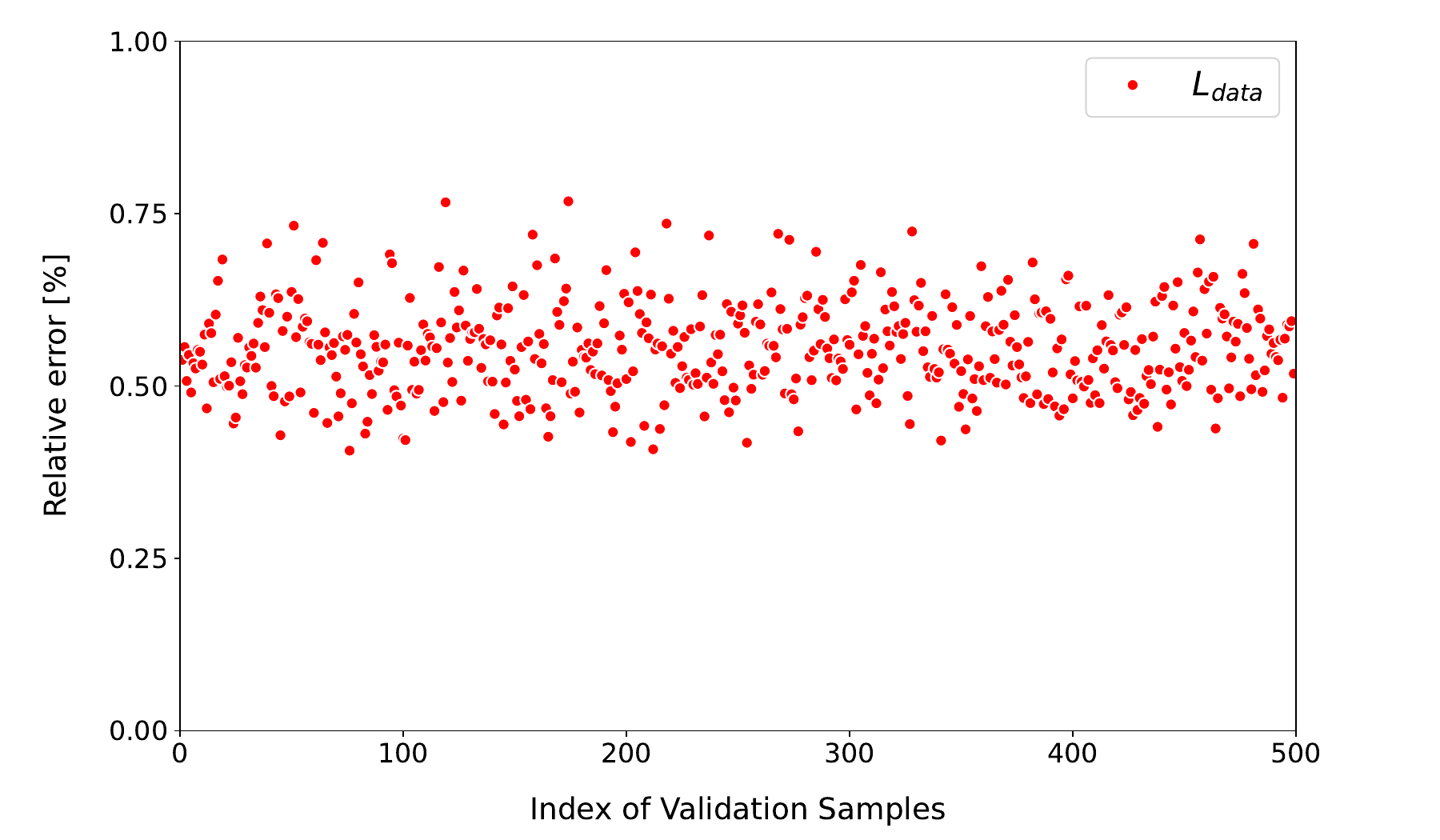}
\caption{The relative error of $\mathcal{L}_{data}$ for the validation set.
The horizontal axis represents the index of the samples, 
while the vertical axis is the relative error.}
\label{fig.testing}
\end{figure}

% validation test
To test our model, we present the validation test by showing the relative error of $\mathcal{L}_{data}$ for each sample in the validation set in Fig. {\ref{fig.testing}}. 
For all samples, the relative error is around $0.5 \%$.
This error can be further reduced by using more samples in the data and physics sets, extending the training to longer epochs, and increasing the size of the NQP model. 
It should be pointed out that this error corresponds to the overall deviation of all the entries of $\hat{\rho}_{\vec{n}}$, including those deep hierarchy elements that have much smaller magnitude as compared to $\hat{\rho}_{\vec{0}}$.

% Test on population dynamics
\subsection{Population dynamics}

By using the composition property, $\vec{\rho}_{t_1 + t_2}  = {\bm{G}}_{t_2}[\theta] \vec{\rho}_{t_1}$, our NQP model can infer truly long-time dynamics well beyond the training time limit $t_{max}$. To test the accuracy of the long-time dynamics predicted by the NQP model, we compute population dynamics up to 40$t_{max}$ ($\sim 1.2 \, {\rm{ps}}$). The reference results are obtained from the RK4 method with an integration time step of $\delta_t$. Here, we consider two initial conditions: (a) $\hat{\rho}_{\vec{0}}(0) = |1\rangle \langle1|$, and $(b)$ $\hat{\rho}_{\vec{0}}(0) = |6 \rangle \langle 6|$, which correspond to the excitation localized at the first and sixth pigment, respectively. All other hierarchy elements $\hat{\rho}_{\vec{n}}(0)$ ($\vec{n}\neq\vec{0}$) are set to zero for the factorized bath initial condition.

% Population
\begin{figure}
\centering
\includegraphics[width=0.45\textwidth]{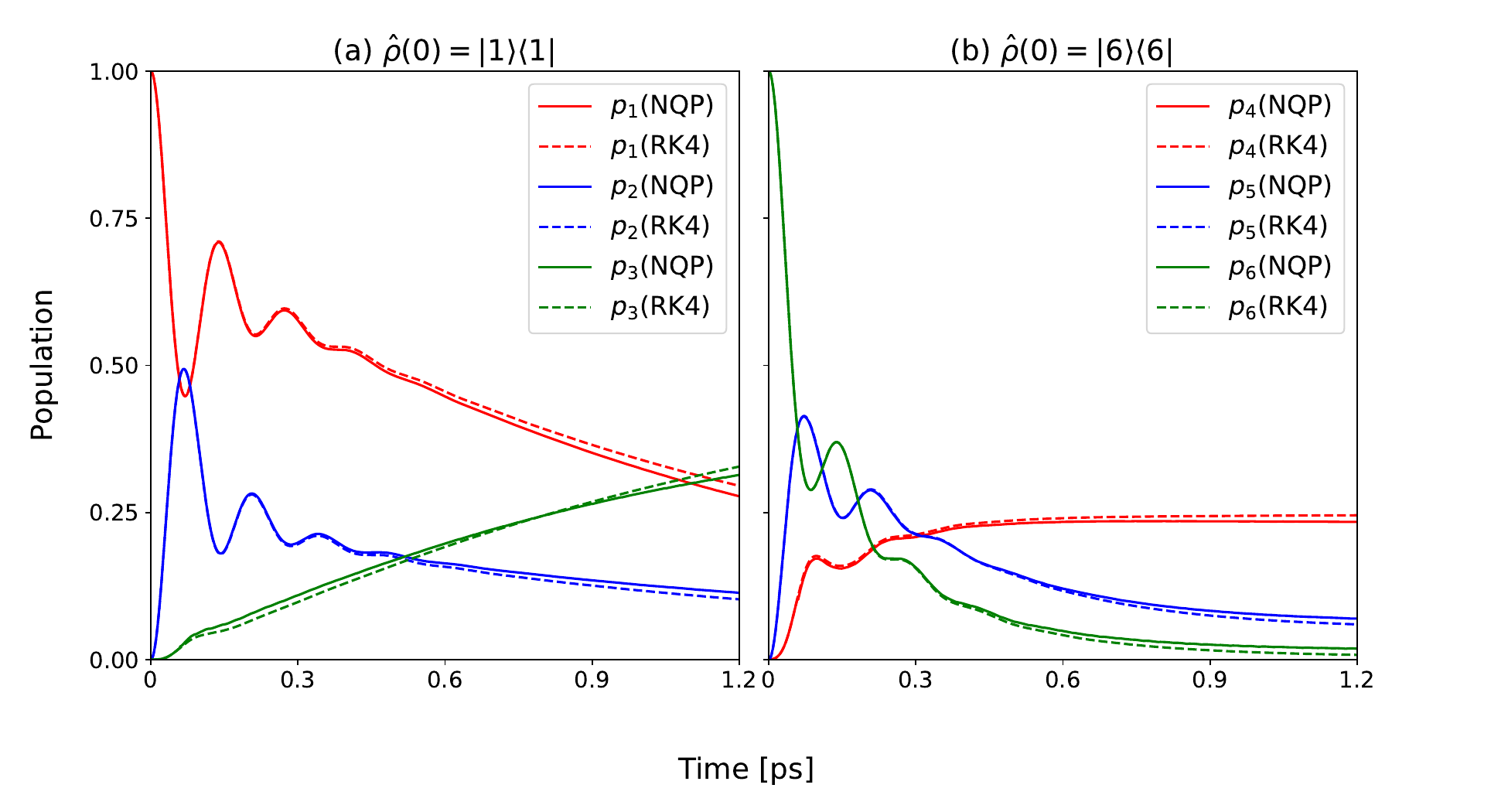}
\caption{Population dynamics computed using NQP model (solid lines) and 
reference RK4 method (dashed lines) for two typical initial conditions up to $t = 1.2 \, {\rm{ps}}$ $(40 \, t_{max})$.}
\label{fig.evolve}
\end{figure}

In Fig. {\ref{fig.evolve}}, we show the time evolution of populations $p_{n}(t) = \langle n | \hat{\rho}_{\vec{0}}(t) | n\rangle$ for sites (a) $n = 1 $, $2$, and $3$, and (b) $n =4$, $5$, and $6$, respectively, following the experimentally demonstrated energy transfer pathways. 
In both cases, our NQP model yields results in perfect agreement with those from the reference RK4 method up to $10 t_{max}$. 
While model-predicted long time dynamics deviates slightly from the exact results due to the accumulation of errors in the training stage, our NQP model still infers the accurate dynamics even far beyond the training time ($40 t_{max}$).

% linear response function
\subsection{Linear spectra}

Next, we apply our NQP model to simulate the linear and third-order response functions as defined in Eqs. {\eqref{eq.resp_linear}} and {\eqref{eq.resp_3rd}}. We choose the transition dipole operator as 
\begin{equation}
\hat{\mu} = \sum_{j=1}^{7} \mu_{j} \left( |j \rangle\langle g| + | g \rangle \langle j| \right),
\end{equation}
where $\mu_{j}$ is the transition dipole moment of $j$-th pigment. The system is initially in the electronic ground state before the photoexcitation, i.e., $\hat{\rho}_{\vec{0}}(0) = |g\rangle \langle g|$.

% linear absorption
\begin{figure}
\centering
\includegraphics[width=0.5\textwidth]{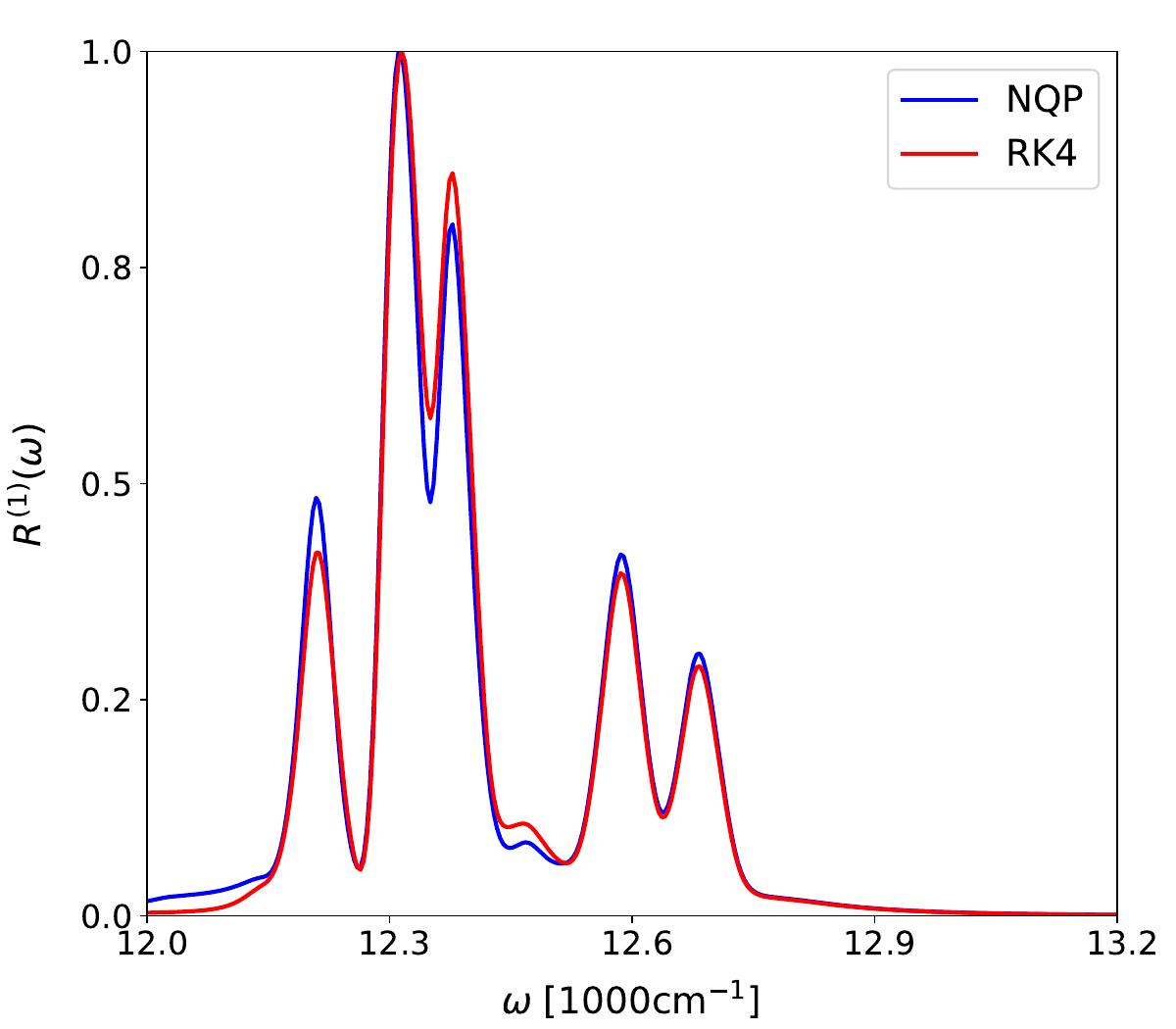}
\caption{The linear spectrum $R^{(1)}(\omega)$ evaluated from the
NQP model (blue line) and the RK4 method (red line), respectively.}
\label{fig.resp_linear}
\end{figure}

In Fig. {\ref{fig.resp_linear}}, we show the linear spectrum evaluated from the NQP model and the RK4 method. For each case, we set the time window of $t$ to $[0, 40 t_{\mathrm{max}}]$,
and normalize the peak intensities with respect to their own maximum value. Overall, the NQP model yields spectrum in good agreement with that from the reference RK4. 
The small deviations of some peak intensities may be attributed to the model's architecture. 
The adaption of Fourier transform in the model's architecture generates some  artificial aliasing modes, the magnitudes of which are increased after recurrent evaluation of long time dynamics. 
This systematic error could be resolved by carefully finetuning the truncation level of Fourier modes in NQP model, or by replacing the Fourier transform with methods such as wavelet transform or spatial convolutions.

\subsection{Two-dimensional spectra}
\label{sec.2des}

% 3rd-order response function
\begin{figure}
\centering
\includegraphics[width=0.5\textwidth]{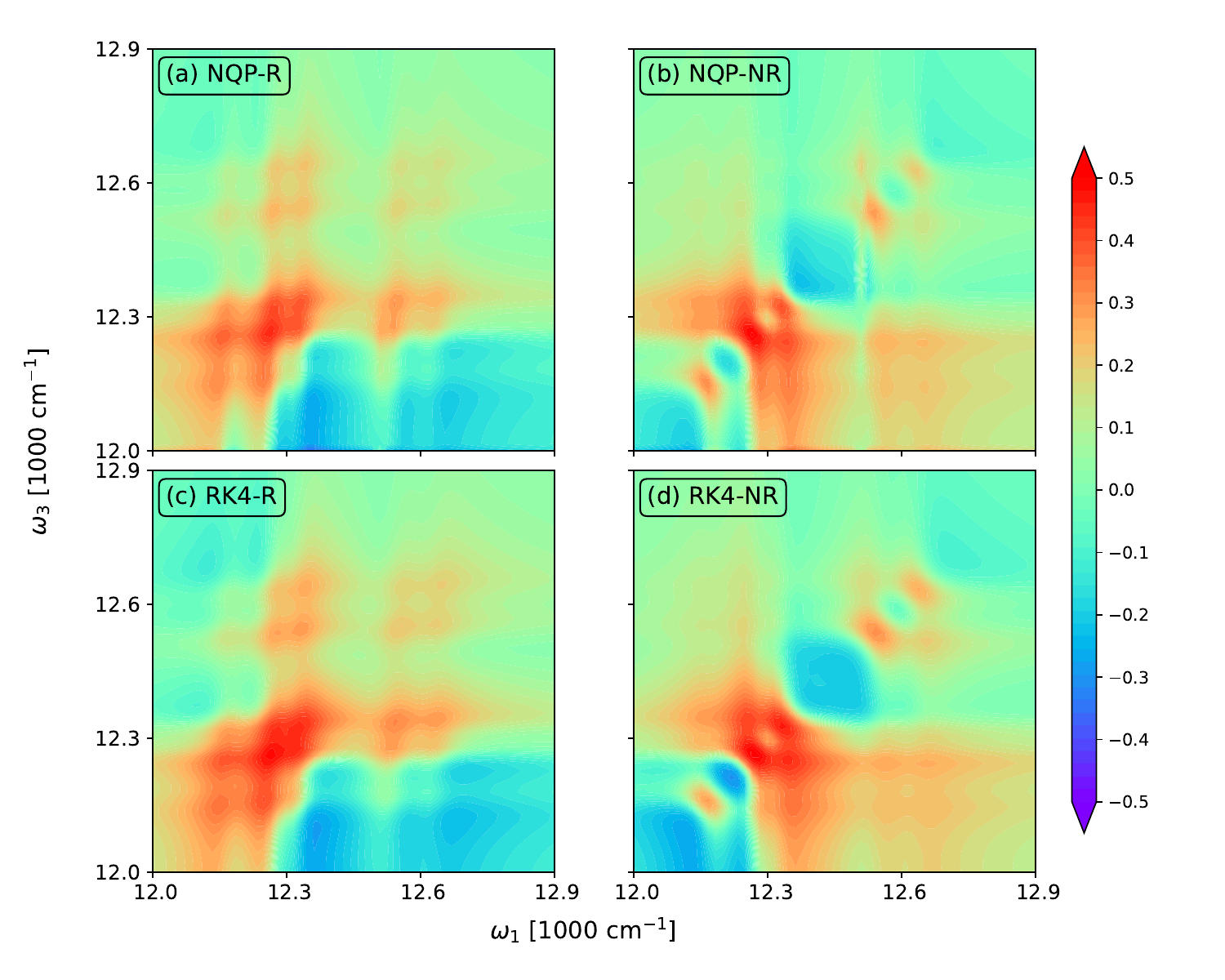}
\caption{2D spectra at $t_{2} = 0 \, {\rm{fs}}$ evaluated from (a,b) the NQP model and (c,d) the RK4 reference, respectively. 
`R' in (a,c) and `NR' in (b,d) denote the rephasing and non-rephasing part, respectively.
}
\label{fig.resp_3rd_0}
\end{figure} 

We further apply the NQP model to compute 2D spectra at different time $t_2$. Since we restrict the discussion to the one-exciton manifold, the two-exciton states, which contribute to the excited-state absorption of 2D spectra, are not included in the Hamiltonian. The simulated 2D spectra thus only involve the contribution of ground-state bleach and stimulated emission. The inclusion of two-exciton states can be easily established by adding the electronic states $|j j^{\prime}\rangle$,
which represent the simultaneous excitation at two pigments, to the Hamiltonian, 
and increasing the length of column vector $\vec{\rho}_{t}$ defined in Eq. {\eqref{eq.heom_adv}}.
The NQP model can handle this expanded system by using more layers and increasing the size of each layer.
However, this will require more learnable parameters and lead to a dramatic increase of computational cost especially in the training stage, which is outside the scope of the present work.

\begin{figure}
\centering
\includegraphics[width=0.45\textwidth]{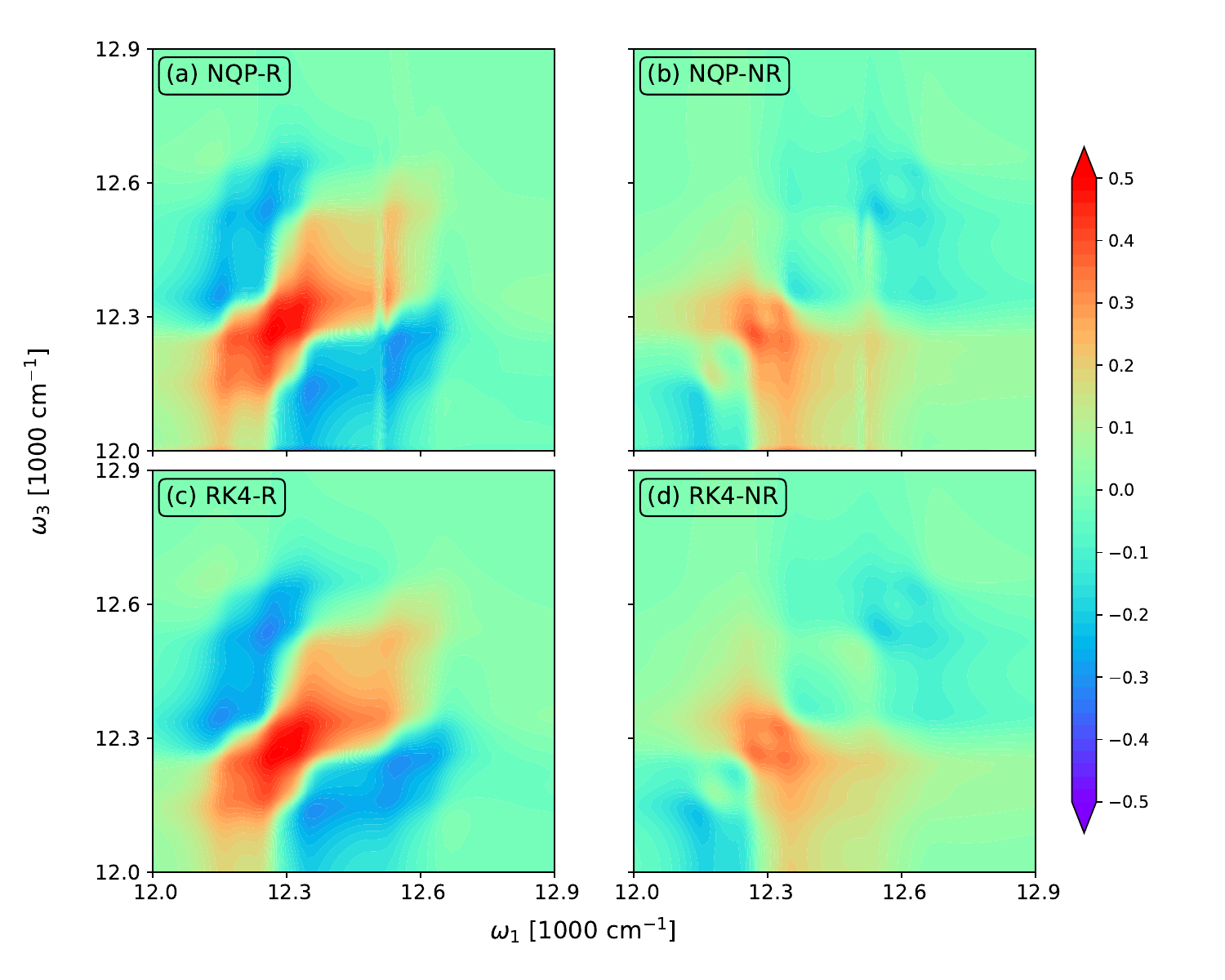}
\caption{The rephasing (a, c) and non-rephasing (b, d) parts of 2D spectra  
at $t_{2} = 50 \, {\rm{fs}}$ evaluated from 
(a, b) the NQP model and (c, d) the RK4 reference, respectively.
}
\label{fig.resp_3rd_50}
\end{figure}

\begin{figure}
\centering
\includegraphics[width=0.45\textwidth]{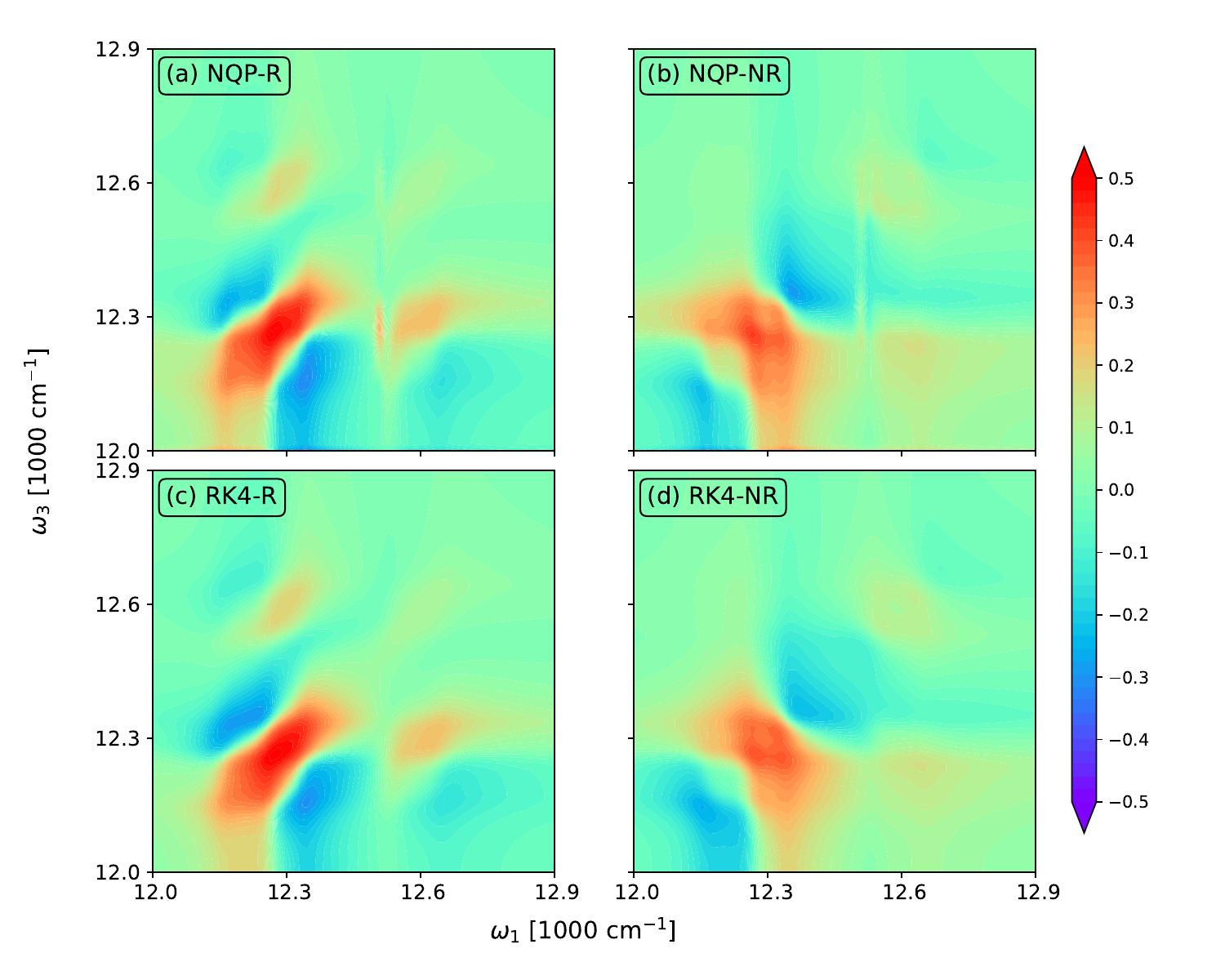}
\caption{
The rephasing (a, c) and non-rephasing (b, d) parts of 2D spectra  
at $t_{2} = 100 \, {\rm{fs}}$ evaluated from 
(a, b) the NQP model and (c, d) the RK4 reference, respectively.
}
\label{fig.resp_3rd_100}
\end{figure}

\begin{figure}
\centering
\includegraphics[width=0.45\textwidth]{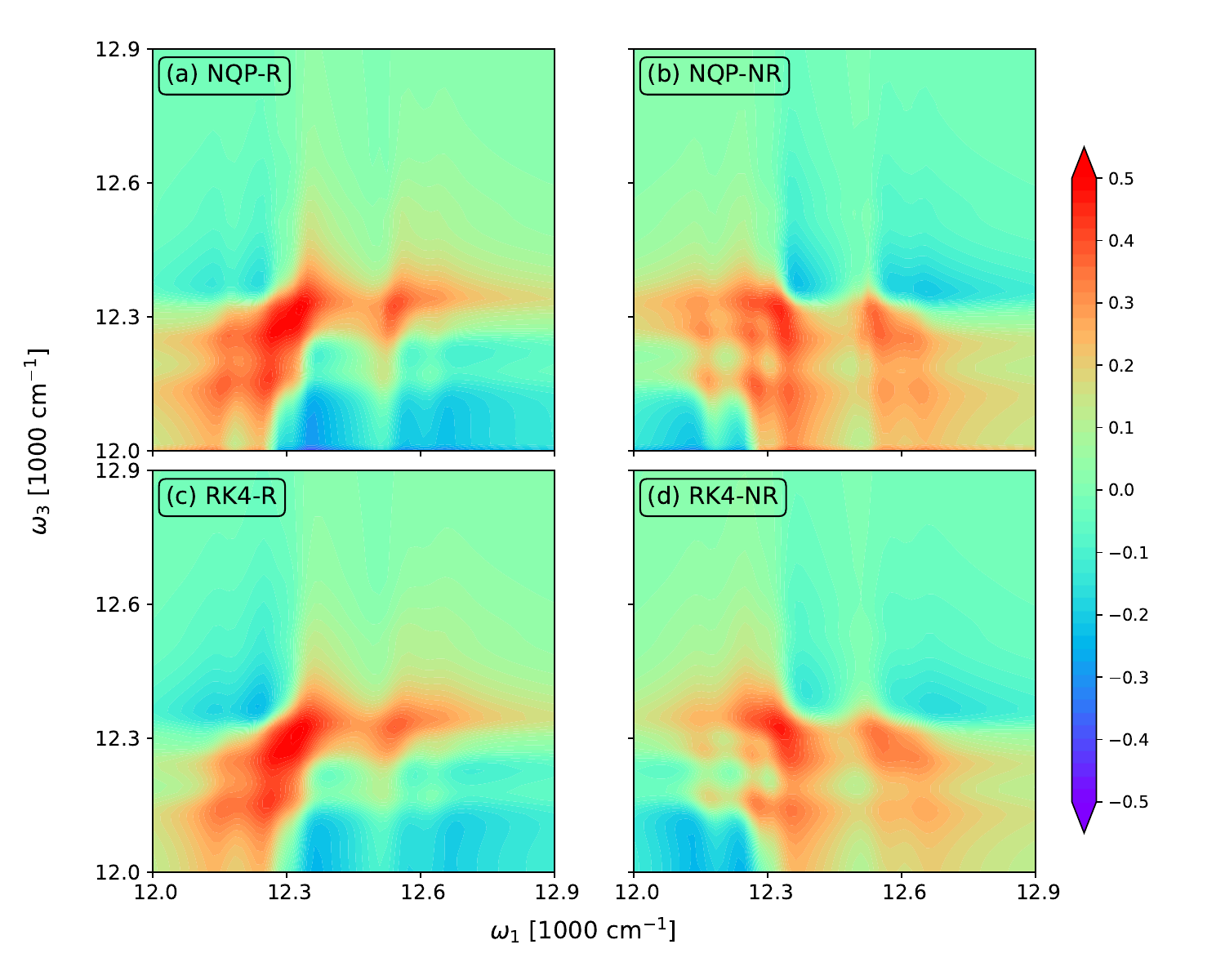}
\caption{
The rephasing (a, c) and non-rephasing (b, d) parts of 2D spectra  
at $t_{2} = 500 \, {\rm{fs}}$ evaluated from 
(a, b) the NQP model and (c, d) the RK4 reference, respectively.
}
\label{fig.resp_3rd_500}
\end{figure}

In Fig. {\ref{fig.resp_3rd_0}}, we show the rephasing (a, c) and non-rephasing (b, d) parts of 2D spectra 
at $t_{2} = 0$ fs evaluated from the NQP model (a, b) and the RK4 reference (c, d), respectively. We set the time window of both $t_1$ and $t_3$ to  $[0, 40 \, t_{\mathrm{max}}]$. 
For better illustration, we adopt the widely used arcsinh scaling after the normalization of peak intensities. {\cite{yeh2014jcp, chen2011jcp, cho2005jpcb}} While the simulation of 2D spectra at $t_2 = 0$ fs requires the propagation up to $t_1 + t_3 \le 80 t_{\mathrm{max}}$, far beyond the training time limit, both rephasing and non-rephasing parts of 2D spectra predicted by the NQP model are again in good agreement with those from the RK4 reference, demonstrating the long-time reliability of our NQP model. The rephasing and non-rephasing parts of 2D spectra at $t_2 = 50,100,500$ fs evaluated from the NQP model and the RK4 reference are shown in Figs.~\ref{fig.resp_3rd_50}-\ref{fig.resp_3rd_500}. The NQP model again yields accurate 2D spectra, illustrating the energy relaxation process from higher exciton states to lower exciton states as $t_2$ increases.

\begin{figure}
\centering
\includegraphics[width=0.5\textwidth]{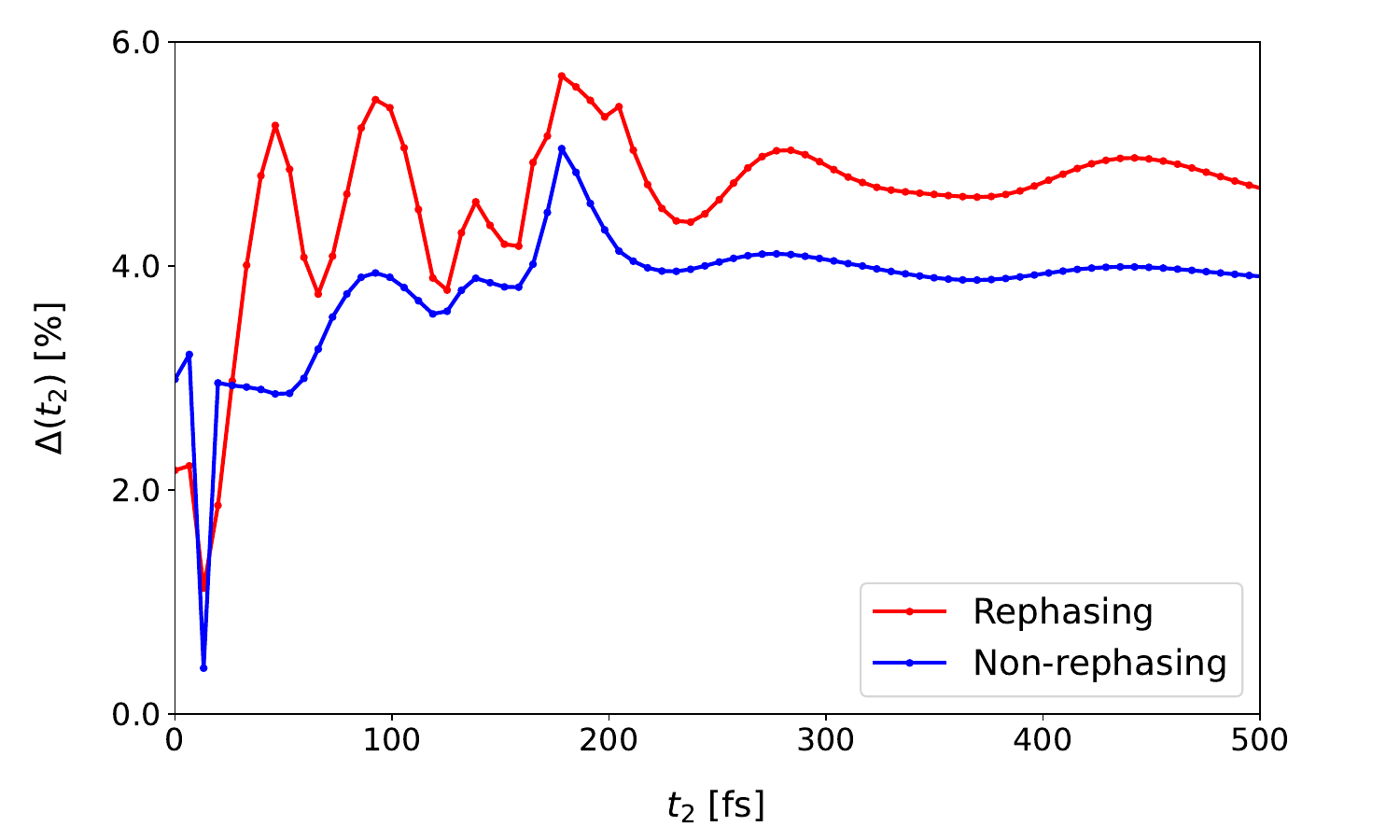}
\caption{The relative error between the NQP model and the RK4 reference as a function of $t_{2}$.}
\label{fig.resp_3rd_diff}
\end{figure}

To quantify the difference between model-predicted and reference spectra more specifically, we define the averaged point-wise relative deviation as
\begin{equation}
\Delta (t_2) = \int_{\Omega_3} {\rm{d}}\omega_3  \int_{\Omega_1} {\rm{d}} 
\omega_1 \,  \left| 1 - \frac{R_{NQP}^{(3)}(\omega_3, \omega_1; t_2) + \varepsilon}
{R_{RK4}^{(3)}(\omega_3, \omega_1; t_2) + \varepsilon} \right| ,
\end{equation}
where $\Omega_1, \Omega_3 \in (12000 {\rm{cm}}^{-1}, 12900 {\rm{cm}}^{-1})$,
$R_{NQP}^{(3)}$ and $R_{RK4}^{(3)}$ are the spectra evaluated from the NQP model and the RK4 reference, respectively, and $\varepsilon \sim 10^{-6}$ is employed to avoid the numerical divergence. In Fig. {\ref{fig.resp_3rd_diff}}, we display $\Delta (t_2)$ up to $t_2 = 500 \, {\rm{fs}}$ with a time step of $20 \, {\rm{fs}}$. 
With the increase of $t_2$, the relative error also rapidly increases due to the fast relaxation of the system. 
After the system reaches the equilibrium, the error also reaches the plateau at around $4\% \sim 5\%$. 
This again demonstrates the long-time reliability of our NQP model.

% 3rd-order response function, difference

%%%%%%%%%%%%
% conclusion
\section{Conclusion}
\label{sec.conclusion}

In this work, we develop a NQP model for the HEOM approach to treat the non-markovian dynamics. 
We use the FNO as the model's architecture, and design a super-resolution algorithm to reduce the computational cost in the data preparation stage. 
In the training stage, we employ both data loss function and an extra PILF to improve the numerical performance. 
The accuracy of the NQP model is tested by computing the population dynamics and linear and two-dimensional spectra. 
The NQP model yields results in good agreement with the conventional RK4 method, demonstrating its potential applicability in various computational scenarios.

In the current NQP model, the number of learnable parameters scales exponentially with both the size of the reduced system and the number of hierarchy elements. 
To alleviate the computational cost, future work may extend the NQP model to deal with the decomposed form of $\hat{\rho}_{\vec{n}}(t)$. For example, $\hat{\rho}_{\vec{n}}(t)$ can be expressed as a matrix-product-state form by using the twin-space formulation and tensor-train decomposition, which requires far less entries than the original density matrix.{\cite{borrelli2019jcp, borrelli2021wirescms}}
In addition, our focus here is on the time-independent Hamiltonian. Future development may extend to the driven dynamics, where the system Hamiltonian contains time-dependent external fields. 
A typical application is the spectroscopic equations of motion approach which was employed to the simulation of strong-field nonlinear spectra. {\cite{wang2004chemphyslett, wang2008chemphys}} 
Work in these directions is in progress.

%%%%%%%%%%%%
% acknowledgement
%%%%%%%%%%%%%
\section*{Acknowledgments}

J.Z. and L.P.C. acknowledge support from the starting grant of research center of new materials computing of Zhejiang Lab (No. 3700-32601).

%%%%%%%%%%%%%
\section*{Author declarations}

\subsection*{Author Contributions}
{\textbf{Jiaji Zhang}}: Data curation (lead); Formal analysis (lead); 
Investigation (equal); Supervision (equal). 

{\textbf{Lipeng Chen}}: Conceptualization (lead); Funding acquisition (lead); 
Investigation (equal); Supervision (equal).

\subsection*{Conflict of Interest}
The authors have no conflicts to disclose.

\section*{Data availability}
The data that support the findings of this study are available from 
the corresponding author upon reasonable request.

\section*{Code availability}
The code that support the findings of this study are available from 
the corresponding author upon reasonable request.

%%%%%%%%%%%%
% bibliography
\bibliography{ref_ls}

%aipnum4-2.bst 2019-01-14 (MD) hand-edited version of apsrev4-1.bst
\begin{thebibliography}{53}%
\makeatletter
\providecommand \@ifxundefined [1]{%
 \@ifx{#1\undefined}
}%
\providecommand \@ifnum [1]{%
 \ifnum #1\expandafter \@firstoftwo
 \else \expandafter \@secondoftwo
 \fi
}%
\providecommand \@ifx [1]{%
 \ifx #1\expandafter \@firstoftwo
 \else \expandafter \@secondoftwo
 \fi
}%
\providecommand \natexlab [1]{#1}%
\providecommand \enquote  [1]{``#1''}%
\providecommand \bibnamefont  [1]{#1}%
\providecommand \bibfnamefont [1]{#1}%
\providecommand \citenamefont [1]{#1}%
\providecommand \href@noop [0]{\@secondoftwo}%
\providecommand \href [0]{\begingroup \@sanitize@url \@href}%
\providecommand \@href[1]{\@@startlink{#1}\@@href}%
\providecommand \@@href[1]{\endgroup#1\@@endlink}%
\providecommand \@sanitize@url [0]{\catcode `\\12\catcode `\$12\catcode
  `\&12\catcode `\#12\catcode `\^12\catcode `\_12\catcode `\%12\relax}%
\providecommand \@@startlink[1]{}%
\providecommand \@@endlink[0]{}%
\providecommand \url  [0]{\begingroup\@sanitize@url \@url }%
\providecommand \@url [1]{\endgroup\@href {#1}{\urlprefix }}%
\providecommand \urlprefix  [0]{URL }%
\providecommand \Eprint [0]{\href }%
\providecommand \doibase [0]{https://doi.org/}%
\providecommand \selectlanguage [0]{\@gobble}%
\providecommand \bibinfo  [0]{\@secondoftwo}%
\providecommand \bibfield  [0]{\@secondoftwo}%
\providecommand \translation [1]{[#1]}%
\providecommand \BibitemOpen [0]{}%
\providecommand \bibitemStop [0]{}%
\providecommand \bibitemNoStop [0]{.\EOS\space}%
\providecommand \EOS [0]{\spacefactor3000\relax}%
\providecommand \BibitemShut  [1]{\csname bibitem#1\endcsname}%
\let\auto@bib@innerbib\@empty
%</preamble>
\bibitem [{\citenamefont {Gelin}, \citenamefont {Chen},\ and\ \citenamefont
  {Domcke}(2022)}]{gelin2022chemrev}%
  \BibitemOpen
  \bibfield  {author} {\bibinfo {author} {\bibfnamefont {M.~F.}\ \bibnamefont
  {Gelin}}, \bibinfo {author} {\bibfnamefont {L.}~\bibnamefont {Chen}},\ and\
  \bibinfo {author} {\bibfnamefont {W.}~\bibnamefont {Domcke}},\ }\href
  {https://doi.org/10.1021/acs.chemrev.2c00329} {\bibfield  {journal} {\bibinfo
   {journal} {Chem. Rev.}\ }\textbf {\bibinfo {volume} {122}},\ \bibinfo
  {pages} {17339--17396} (\bibinfo {year} {2022})}\BibitemShut {NoStop}%
\bibitem [{\citenamefont {Nisoli}\ \emph {et~al.}(2017)\citenamefont {Nisoli},
  \citenamefont {Decleva}, \citenamefont {Calegari}, \citenamefont {Palacios},\
  and\ \citenamefont {Martín}}]{nisoli2017chemrev}%
  \BibitemOpen
  \bibfield  {author} {\bibinfo {author} {\bibfnamefont {M.}~\bibnamefont
  {Nisoli}}, \bibinfo {author} {\bibfnamefont {P.}~\bibnamefont {Decleva}},
  \bibinfo {author} {\bibfnamefont {F.}~\bibnamefont {Calegari}}, \bibinfo
  {author} {\bibfnamefont {A.}~\bibnamefont {Palacios}},\ and\ \bibinfo
  {author} {\bibfnamefont {F.}~\bibnamefont {Martín}},\ }\href
  {https://doi.org/10.1021/acs.chemrev.6b00453} {\bibfield  {journal} {\bibinfo
   {journal} {Chem. Rev.}\ }\textbf {\bibinfo {volume} {117}},\ \bibinfo
  {pages} {10760--10825} (\bibinfo {year} {2017})}\BibitemShut {NoStop}%
\bibitem [{\citenamefont {Mukamel}(2000)}]{mukamel2000arpc}%
  \BibitemOpen
  \bibfield  {author} {\bibinfo {author} {\bibfnamefont {S.}~\bibnamefont
  {Mukamel}},\ }\href {https://doi.org/10.1146/annurev.physchem.51.1.691}
  {\bibfield  {journal} {\bibinfo  {journal} {Annu. Rev. Phys. Chem.}\ }\textbf
  {\bibinfo {volume} {51}},\ \bibinfo {pages} {691–729} (\bibinfo {year}
  {2000})}\BibitemShut {NoStop}%
\bibitem [{\citenamefont {Maiuri}, \citenamefont {Garavelli},\ and\
  \citenamefont {Cerullo}(2019)}]{maiuri2019jacs}%
  \BibitemOpen
  \bibfield  {author} {\bibinfo {author} {\bibfnamefont {M.}~\bibnamefont
  {Maiuri}}, \bibinfo {author} {\bibfnamefont {M.}~\bibnamefont {Garavelli}},\
  and\ \bibinfo {author} {\bibfnamefont {G.}~\bibnamefont {Cerullo}},\ }\href
  {https://doi.org/10.1021/jacs.9b10533} {\bibfield  {journal} {\bibinfo
  {journal} {J. Am. Chem. Soc.}\ }\textbf {\bibinfo {volume} {142}},\ \bibinfo
  {pages} {3–15} (\bibinfo {year} {2019})}\BibitemShut {NoStop}%
\bibitem [{\citenamefont {Dorfman}, \citenamefont {Schlawin},\ and\
  \citenamefont {Mukamel}(2016)}]{dorfman2016rmv}%
  \BibitemOpen
  \bibfield  {author} {\bibinfo {author} {\bibfnamefont {K.~E.}\ \bibnamefont
  {Dorfman}}, \bibinfo {author} {\bibfnamefont {F.}~\bibnamefont {Schlawin}},\
  and\ \bibinfo {author} {\bibfnamefont {S.}~\bibnamefont {Mukamel}},\ }\href
  {https://doi.org/10.1103/RevModPhys.88.045008} {\bibfield  {journal}
  {\bibinfo  {journal} {Rev. Mod. Phys.}\ }\textbf {\bibinfo {volume} {88}},\
  \bibinfo {pages} {045008} (\bibinfo {year} {2016})}\BibitemShut {NoStop}%
\bibitem [{\citenamefont {Fresch}\ \emph {et~al.}(2023)\citenamefont {Fresch},
  \citenamefont {Camargo}, \citenamefont {Shen}, \citenamefont {Bellora},
  \citenamefont {Pullerits}, \citenamefont {Engel}, \citenamefont {Cerullo},\
  and\ \citenamefont {Collini}}]{fresch2023nrmp}%
  \BibitemOpen
  \bibfield  {author} {\bibinfo {author} {\bibfnamefont {E.}~\bibnamefont
  {Fresch}}, \bibinfo {author} {\bibfnamefont {F.~V.~A.}\ \bibnamefont
  {Camargo}}, \bibinfo {author} {\bibfnamefont {Q.}~\bibnamefont {Shen}},
  \bibinfo {author} {\bibfnamefont {C.~C.}\ \bibnamefont {Bellora}}, \bibinfo
  {author} {\bibfnamefont {T.}~\bibnamefont {Pullerits}}, \bibinfo {author}
  {\bibfnamefont {G.~S.}\ \bibnamefont {Engel}}, \bibinfo {author}
  {\bibfnamefont {G.}~\bibnamefont {Cerullo}},\ and\ \bibinfo {author}
  {\bibfnamefont {E.}~\bibnamefont {Collini}},\ }\href@noop {} {\bibfield
  {journal} {\bibinfo  {journal} {Nat. Rev. Methods Primers}\ }\textbf
  {\bibinfo {volume} {3}} (\bibinfo {year} {2023})}\BibitemShut {NoStop}%
\bibitem [{\citenamefont {Oliver}(2018)}]{oliver2018rsos}%
  \BibitemOpen
  \bibfield  {author} {\bibinfo {author} {\bibfnamefont {T.~A.~A.}\
  \bibnamefont {Oliver}},\ }\href {https://doi.org/10.1098/rsos.171425}
  {\bibfield  {journal} {\bibinfo  {journal} {R. Soc. Open Sci.}\ }\textbf
  {\bibinfo {volume} {5}},\ \bibinfo {pages} {171425} (\bibinfo {year}
  {2018})}\BibitemShut {NoStop}%
\bibitem [{\citenamefont {Schlau-Cohen}, \citenamefont {Ishizaki},\ and\
  \citenamefont {Fleming}(2011)}]{cohen2011chemphys}%
  \BibitemOpen
  \bibfield  {author} {\bibinfo {author} {\bibfnamefont {G.~S.}\ \bibnamefont
  {Schlau-Cohen}}, \bibinfo {author} {\bibfnamefont {A.}~\bibnamefont
  {Ishizaki}},\ and\ \bibinfo {author} {\bibfnamefont {G.~R.}\ \bibnamefont
  {Fleming}},\ }\href
  {https://doi.org/https://doi.org/10.1016/j.chemphys.2011.04.025} {\bibfield
  {journal} {\bibinfo  {journal} {Chem. Phys.}\ }\textbf {\bibinfo {volume}
  {386}},\ \bibinfo {pages} {1--22} (\bibinfo {year} {2011})}\BibitemShut
  {NoStop}%
\bibitem [{\citenamefont {Ginsberg}, \citenamefont {Cheng},\ and\ \citenamefont
  {Fleming}(2009)}]{ginsberg2009acr}%
  \BibitemOpen
  \bibfield  {author} {\bibinfo {author} {\bibfnamefont {N.~S.}\ \bibnamefont
  {Ginsberg}}, \bibinfo {author} {\bibfnamefont {Y.-C.}\ \bibnamefont
  {Cheng}},\ and\ \bibinfo {author} {\bibfnamefont {G.~R.}\ \bibnamefont
  {Fleming}},\ }\href {https://doi.org/10.1021/ar9001075} {\bibfield  {journal}
  {\bibinfo  {journal} {Acc. Chem. Res.}\ }\textbf {\bibinfo {volume} {42}},\
  \bibinfo {pages} {1352–1363} (\bibinfo {year} {2009})}\BibitemShut
  {NoStop}%
\bibitem [{\citenamefont {Scholes}\ \emph {et~al.}(2011)\citenamefont
  {Scholes}, \citenamefont {Fleming}, \citenamefont {Olaya-Castro},\ and\
  \citenamefont {van Grondelle}}]{scholes2011natchem}%
  \BibitemOpen
  \bibfield  {author} {\bibinfo {author} {\bibfnamefont {G.~D.}\ \bibnamefont
  {Scholes}}, \bibinfo {author} {\bibfnamefont {G.~R.}\ \bibnamefont
  {Fleming}}, \bibinfo {author} {\bibfnamefont {A.}~\bibnamefont
  {Olaya-Castro}},\ and\ \bibinfo {author} {\bibfnamefont {R.}~\bibnamefont
  {van Grondelle}},\ }\href {https://doi.org/10.1038/nchem.1145} {\bibfield
  {journal} {\bibinfo  {journal} {Nat. Chem.}\ }\textbf {\bibinfo {volume}
  {3}},\ \bibinfo {pages} {763–774} (\bibinfo {year} {2011})}\BibitemShut
  {NoStop}%
\bibitem [{\citenamefont {Kullmann}\ \emph {et~al.}(2011)\citenamefont
  {Kullmann}, \citenamefont {Ruetzel}, \citenamefont {Buback}, \citenamefont
  {Nuernberger},\ and\ \citenamefont {Brixner}}]{martin2011jacs}%
  \BibitemOpen
  \bibfield  {author} {\bibinfo {author} {\bibfnamefont {M.}~\bibnamefont
  {Kullmann}}, \bibinfo {author} {\bibfnamefont {S.}~\bibnamefont {Ruetzel}},
  \bibinfo {author} {\bibfnamefont {J.}~\bibnamefont {Buback}}, \bibinfo
  {author} {\bibfnamefont {P.}~\bibnamefont {Nuernberger}},\ and\ \bibinfo
  {author} {\bibfnamefont {T.}~\bibnamefont {Brixner}},\ }\href
  {https://doi.org/10.1021/ja2032037} {\bibfield  {journal} {\bibinfo
  {journal} {J. Am. Chem. Soc.}\ }\textbf {\bibinfo {volume} {133}},\ \bibinfo
  {pages} {13074--13080} (\bibinfo {year} {2011})}\BibitemShut {NoStop}%
\bibitem [{\citenamefont {Arsenault}\ \emph {et~al.}(2021)\citenamefont
  {Arsenault}, \citenamefont {Bhattacharyya}, \citenamefont {Yoneda},\ and\
  \citenamefont {Fleming}}]{arsenault2021jcp}%
  \BibitemOpen
  \bibfield  {author} {\bibinfo {author} {\bibfnamefont {E.~A.}\ \bibnamefont
  {Arsenault}}, \bibinfo {author} {\bibfnamefont {P.}~\bibnamefont
  {Bhattacharyya}}, \bibinfo {author} {\bibfnamefont {Y.}~\bibnamefont
  {Yoneda}},\ and\ \bibinfo {author} {\bibfnamefont {G.~R.}\ \bibnamefont
  {Fleming}},\ }\href@noop {} {\bibfield  {journal} {\bibinfo  {journal} {J.
  Chem. Phys.}\ }\textbf {\bibinfo {volume} {155}} (\bibinfo {year}
  {2021})}\BibitemShut {NoStop}%
\bibitem [{\citenamefont {Kim}\ \emph {et~al.}(2020)\citenamefont {Kim},
  \citenamefont {Jeon}, \citenamefont {Yoon},\ and\ \citenamefont
  {Cho}}]{kim2020natcomm}%
  \BibitemOpen
  \bibfield  {author} {\bibinfo {author} {\bibfnamefont {J.}~\bibnamefont
  {Kim}}, \bibinfo {author} {\bibfnamefont {J.}~\bibnamefont {Jeon}}, \bibinfo
  {author} {\bibfnamefont {T.~H.}\ \bibnamefont {Yoon}},\ and\ \bibinfo
  {author} {\bibfnamefont {M.}~\bibnamefont {Cho}},\ }\href@noop {} {\bibfield
  {journal} {\bibinfo  {journal} {Nat. Comm.}\ }\textbf {\bibinfo {volume}
  {11}} (\bibinfo {year} {2020})}\BibitemShut {NoStop}%
\bibitem [{\citenamefont {Ruetzel}\ \emph {et~al.}(2013)\citenamefont
  {Ruetzel}, \citenamefont {Kullmann}, \citenamefont {Buback}, \citenamefont
  {Nuernberger},\ and\ \citenamefont {Brixner}}]{ruetzel2013prl}%
  \BibitemOpen
  \bibfield  {author} {\bibinfo {author} {\bibfnamefont {S.}~\bibnamefont
  {Ruetzel}}, \bibinfo {author} {\bibfnamefont {M.}~\bibnamefont {Kullmann}},
  \bibinfo {author} {\bibfnamefont {J.}~\bibnamefont {Buback}}, \bibinfo
  {author} {\bibfnamefont {P.}~\bibnamefont {Nuernberger}},\ and\ \bibinfo
  {author} {\bibfnamefont {T.}~\bibnamefont {Brixner}},\ }\href@noop {}
  {\bibfield  {journal} {\bibinfo  {journal} {Phys. Rev. Lett}\ }\textbf
  {\bibinfo {volume} {110}} (\bibinfo {year} {2013})}\BibitemShut {NoStop}%
\bibitem [{\citenamefont {Cho}(2019)}]{cho2019}%
  \BibitemOpen
  \bibfield  {author} {\bibinfo {author} {\bibfnamefont {M.}~\bibnamefont
  {Cho}},\ }\href {https://doi.org/10.1007/978-981-13-9753-0} {\emph {\bibinfo
  {title} {Coherent Multidimensional Spectroscopy}}}\ (\bibinfo  {publisher}
  {Springer Singapore},\ \bibinfo {year} {2019})\BibitemShut {NoStop}%
\bibitem [{\citenamefont {Mukamel}(1995)}]{mukamel1999}%
  \BibitemOpen
  \bibfield  {author} {\bibinfo {author} {\bibfnamefont {S.}~\bibnamefont
  {Mukamel}},\ }\href@noop {} {\emph {\bibinfo {title} {Principles of Nonlinear
  Optical Spectroscopy}}}\ (\bibinfo  {publisher} {Oxford University Press},\
  \bibinfo {year} {1995})\BibitemShut {NoStop}%
\bibitem [{\citenamefont {Breuer}\ and\ \citenamefont
  {Petruccione}(2007)}]{breuer2007}%
  \BibitemOpen
  \bibfield  {author} {\bibinfo {author} {\bibfnamefont {H.-P.}\ \bibnamefont
  {Breuer}}\ and\ \bibinfo {author} {\bibfnamefont {F.}~\bibnamefont
  {Petruccione}},\ }\href@noop {} {\emph {\bibinfo {title} {The Theory of Open
  Quantum Systems}}}\ (\bibinfo  {publisher} {Oxford University Press},\
  \bibinfo {year} {2007})\BibitemShut {NoStop}%
\bibitem [{\citenamefont {Weiss}(2012)}]{weiss2012}%
  \BibitemOpen
  \bibfield  {author} {\bibinfo {author} {\bibfnamefont {U.}~\bibnamefont
  {Weiss}},\ }\href@noop {} {\emph {\bibinfo {title} {Quantum Dissipative
  Systems}}},\ \bibinfo {edition} {4th}\ ed.\ (\bibinfo  {publisher} {World
  Scientific},\ \bibinfo {year} {2012})\BibitemShut {NoStop}%
\bibitem [{\citenamefont {Tanimura}(2020)}]{tanimura2020jcp}%
  \BibitemOpen
  \bibfield  {author} {\bibinfo {author} {\bibfnamefont {Y.}~\bibnamefont
  {Tanimura}},\ }\href {https://doi.org/10.1063/5.0011599} {\bibfield
  {journal} {\bibinfo  {journal} {J. Chem. Phys.}\ }\textbf {\bibinfo {volume}
  {153}},\ \bibinfo {pages} {020901} (\bibinfo {year} {2020})}\BibitemShut
  {NoStop}%
\bibitem [{\citenamefont {Ye}\ \emph {et~al.}(2016)\citenamefont {Ye},
  \citenamefont {Wang}, \citenamefont {Hou}, \citenamefont {Xu}, \citenamefont
  {Zheng},\ and\ \citenamefont {Yan}}]{ye2016wirescms}%
  \BibitemOpen
  \bibfield  {author} {\bibinfo {author} {\bibfnamefont {L.}~\bibnamefont
  {Ye}}, \bibinfo {author} {\bibfnamefont {X.}~\bibnamefont {Wang}}, \bibinfo
  {author} {\bibfnamefont {D.}~\bibnamefont {Hou}}, \bibinfo {author}
  {\bibfnamefont {R.}~\bibnamefont {Xu}}, \bibinfo {author} {\bibfnamefont
  {X.}~\bibnamefont {Zheng}},\ and\ \bibinfo {author} {\bibfnamefont
  {Y.}~\bibnamefont {Yan}},\ }\href {https://doi.org/10.1002/wcms.1269}
  {\bibfield  {journal} {\bibinfo  {journal} {WIREs Comput. Mol. Sci.}\
  }\textbf {\bibinfo {volume} {6}},\ \bibinfo {pages} {608–638} (\bibinfo
  {year} {2016})}\BibitemShut {NoStop}%
\bibitem [{\citenamefont {Zhang}\ and\ \citenamefont
  {Tanimura}(2022)}]{zhang2022imheom}%
  \BibitemOpen
  \bibfield  {author} {\bibinfo {author} {\bibfnamefont {J.}~\bibnamefont
  {Zhang}}\ and\ \bibinfo {author} {\bibfnamefont {Y.}~\bibnamefont
  {Tanimura}},\ }\href@noop {} {\bibfield  {journal} {\bibinfo  {journal} {J.
  Chem. Phys.}\ }\textbf {\bibinfo {volume} {156}} (\bibinfo {year}
  {2022})}\BibitemShut {NoStop}%
\bibitem [{\citenamefont {Kloeden}\ and\ \citenamefont
  {Platen}(1992)}]{kloeden1992}%
  \BibitemOpen
  \bibfield  {author} {\bibinfo {author} {\bibfnamefont {P.~E.}\ \bibnamefont
  {Kloeden}}\ and\ \bibinfo {author} {\bibfnamefont {E.}~\bibnamefont
  {Platen}},\ }\href {https://doi.org/10.1007/978-3-662-12616-5} {\emph
  {\bibinfo {title} {Numerical Solution of Stochastic Differential
  Equations}}}\ (\bibinfo  {publisher} {Springer Berlin Heidelberg},\ \bibinfo
  {year} {1992})\BibitemShut {NoStop}%
\bibitem [{\citenamefont {Yan}\ \emph {et~al.}(2021)\citenamefont {Yan},
  \citenamefont {Xu}, \citenamefont {Li},\ and\ \citenamefont
  {Shi}}]{yan2021jcp}%
  \BibitemOpen
  \bibfield  {author} {\bibinfo {author} {\bibfnamefont {Y.}~\bibnamefont
  {Yan}}, \bibinfo {author} {\bibfnamefont {M.}~\bibnamefont {Xu}}, \bibinfo
  {author} {\bibfnamefont {T.}~\bibnamefont {Li}},\ and\ \bibinfo {author}
  {\bibfnamefont {Q.}~\bibnamefont {Shi}},\ }\href
  {https://doi.org/10.1063/5.0050720} {\bibfield  {journal} {\bibinfo
  {journal} {J. Chem. Phys.}\ }\textbf {\bibinfo {volume} {154}},\ \bibinfo
  {pages} {194104} (\bibinfo {year} {2021})}\BibitemShut {NoStop}%
\bibitem [{\citenamefont {Ke}(2023)}]{ke2023jcp}%
  \BibitemOpen
  \bibfield  {author} {\bibinfo {author} {\bibfnamefont {Y.}~\bibnamefont
  {Ke}},\ }\href@noop {} {\bibfield  {journal} {\bibinfo  {journal} {J. Chem.
  Phys.}\ }\textbf {\bibinfo {volume} {158}} (\bibinfo {year}
  {2023})}\BibitemShut {NoStop}%
\bibitem [{\citenamefont {Kimura}\ and\ \citenamefont
  {Fujihashi}(2014)}]{kimura2014}%
  \BibitemOpen
  \bibfield  {author} {\bibinfo {author} {\bibfnamefont {A.}~\bibnamefont
  {Kimura}}\ and\ \bibinfo {author} {\bibfnamefont {Y.}~\bibnamefont
  {Fujihashi}},\ }\href {https://doi.org/10.1063/1.4901431} {\bibfield
  {journal} {\bibinfo  {journal} {J. Chem. Phys.}\ }\textbf {\bibinfo {volume}
  {141}},\ \bibinfo {pages} {194110} (\bibinfo {year} {2014})}\BibitemShut
  {NoStop}%
\bibitem [{\citenamefont {Schlimgen}\ \emph {et~al.}(2021)\citenamefont
  {Schlimgen}, \citenamefont {Head-Marsden}, \citenamefont {Sager},
  \citenamefont {Narang},\ and\ \citenamefont {Mazziotti}}]{schlimgen2021prl}%
  \BibitemOpen
  \bibfield  {author} {\bibinfo {author} {\bibfnamefont {A.~W.}\ \bibnamefont
  {Schlimgen}}, \bibinfo {author} {\bibfnamefont {K.}~\bibnamefont
  {Head-Marsden}}, \bibinfo {author} {\bibfnamefont {L.~M.}\ \bibnamefont
  {Sager}}, \bibinfo {author} {\bibfnamefont {P.}~\bibnamefont {Narang}},\ and\
  \bibinfo {author} {\bibfnamefont {D.~A.}\ \bibnamefont {Mazziotti}},\ }\href
  {https://doi.org/10.1103/PhysRevLett.127.270503} {\bibfield  {journal}
  {\bibinfo  {journal} {Phys. Rev. Lett}\ }\textbf {\bibinfo {volume} {127}},\
  \bibinfo {pages} {270503} (\bibinfo {year} {2021})}\BibitemShut {NoStop}%
\bibitem [{\citenamefont {Liu}\ \emph {et~al.}(2023)\citenamefont {Liu},
  \citenamefont {Chen}, \citenamefont {Su}, \citenamefont {Wang},\ and\
  \citenamefont {Dou}}]{liu2023jcp}%
  \BibitemOpen
  \bibfield  {author} {\bibinfo {author} {\bibfnamefont {W.}~\bibnamefont
  {Liu}}, \bibinfo {author} {\bibfnamefont {Z.-H.}\ \bibnamefont {Chen}},
  \bibinfo {author} {\bibfnamefont {Y.}~\bibnamefont {Su}}, \bibinfo {author}
  {\bibfnamefont {Y.}~\bibnamefont {Wang}},\ and\ \bibinfo {author}
  {\bibfnamefont {W.}~\bibnamefont {Dou}},\ }\href
  {https://doi.org/10.1063/5.0170512} {\bibfield  {journal} {\bibinfo
  {journal} {J. Chem. Phys.}\ }\textbf {\bibinfo {volume} {159}},\ \bibinfo
  {pages} {144110} (\bibinfo {year} {2023})}\BibitemShut {NoStop}%
\bibitem [{\citenamefont {LeCun}, \citenamefont {Bengio},\ and\ \citenamefont
  {Hinton}(2015)}]{lecun2015nature}%
  \BibitemOpen
  \bibfield  {author} {\bibinfo {author} {\bibfnamefont {Y.}~\bibnamefont
  {LeCun}}, \bibinfo {author} {\bibfnamefont {Y.}~\bibnamefont {Bengio}},\ and\
  \bibinfo {author} {\bibfnamefont {G.}~\bibnamefont {Hinton}},\ }\href
  {https://doi.org/10.1038/nature14539} {\bibfield  {journal} {\bibinfo
  {journal} {Nature}\ }\textbf {\bibinfo {volume} {521}},\ \bibinfo {pages}
  {436--444} (\bibinfo {year} {2015})}\BibitemShut {NoStop}%
\bibitem [{\citenamefont {Hermann}\ \emph {et~al.}(2023)\citenamefont
  {Hermann}, \citenamefont {Spencer}, \citenamefont {Choo}, \citenamefont
  {Mezzacapo}, \citenamefont {Foulkes}, \citenamefont {Pfau}, \citenamefont
  {Carleo},\ and\ \citenamefont {Noé}}]{hermann2023natrevchem}%
  \BibitemOpen
  \bibfield  {author} {\bibinfo {author} {\bibfnamefont {J.}~\bibnamefont
  {Hermann}}, \bibinfo {author} {\bibfnamefont {J.}~\bibnamefont {Spencer}},
  \bibinfo {author} {\bibfnamefont {K.}~\bibnamefont {Choo}}, \bibinfo {author}
  {\bibfnamefont {A.}~\bibnamefont {Mezzacapo}}, \bibinfo {author}
  {\bibfnamefont {W.~M.~C.}\ \bibnamefont {Foulkes}}, \bibinfo {author}
  {\bibfnamefont {D.}~\bibnamefont {Pfau}}, \bibinfo {author} {\bibfnamefont
  {G.}~\bibnamefont {Carleo}},\ and\ \bibinfo {author} {\bibfnamefont
  {F.}~\bibnamefont {Noé}},\ }\href
  {https://doi.org/10.1038/s41570-023-00516-8} {\bibfield  {journal} {\bibinfo
  {journal} {Nat. Rev. Chem.}\ }\textbf {\bibinfo {volume} {7}},\ \bibinfo
  {pages} {692–709} (\bibinfo {year} {2023})}\BibitemShut {NoStop}%
\bibitem [{\citenamefont {Lu}\ \emph {et~al.}(2021{\natexlab{a}})\citenamefont
  {Lu}, \citenamefont {Jin}, \citenamefont {Pang}, \citenamefont {Zhang},\ and\
  \citenamefont {Karniadakis}}]{lu2021don}%
  \BibitemOpen
  \bibfield  {author} {\bibinfo {author} {\bibfnamefont {L.}~\bibnamefont
  {Lu}}, \bibinfo {author} {\bibfnamefont {P.}~\bibnamefont {Jin}}, \bibinfo
  {author} {\bibfnamefont {G.}~\bibnamefont {Pang}}, \bibinfo {author}
  {\bibfnamefont {Z.}~\bibnamefont {Zhang}},\ and\ \bibinfo {author}
  {\bibfnamefont {G.~E.}\ \bibnamefont {Karniadakis}},\ }\href
  {https://doi.org/10.1038/s42256-021-00302-5} {\bibfield  {journal} {\bibinfo
  {journal} {Nat. Mach. Intell.}\ }\textbf {\bibinfo {volume} {3}},\ \bibinfo
  {pages} {218–229} (\bibinfo {year} {2021}{\natexlab{a}})}\BibitemShut
  {NoStop}%
\bibitem [{\citenamefont {Li}\ \emph {et~al.}()\citenamefont {Li},
  \citenamefont {Kovachki}, \citenamefont {Azizzadenesheli}, \citenamefont
  {Liu}, \citenamefont {Bhattacharya}, \citenamefont {Stuart},\ and\
  \citenamefont {Anandkumar}}]{li2021fno}%
  \BibitemOpen
  \bibfield  {author} {\bibinfo {author} {\bibfnamefont {Z.}~\bibnamefont
  {Li}}, \bibinfo {author} {\bibfnamefont {N.}~\bibnamefont {Kovachki}},
  \bibinfo {author} {\bibfnamefont {K.}~\bibnamefont {Azizzadenesheli}},
  \bibinfo {author} {\bibfnamefont {B.}~\bibnamefont {Liu}}, \bibinfo {author}
  {\bibfnamefont {K.}~\bibnamefont {Bhattacharya}}, \bibinfo {author}
  {\bibfnamefont {A.}~\bibnamefont {Stuart}},\ and\ \bibinfo {author}
  {\bibfnamefont {A.}~\bibnamefont {Anandkumar}},\ }\href@noop {} {\ }\bibinfo
  {note} {Eprint arXiv: 2010.08895 (2021)}\BibitemShut {NoStop}%
\bibitem [{\citenamefont {Kovachki}\ \emph {et~al.}(2023)\citenamefont
  {Kovachki}, \citenamefont {Li}, \citenamefont {Liu}, \citenamefont
  {Azizzadenesheli}, \citenamefont {Bhattacharya}, \citenamefont {Stuart},\
  and\ \citenamefont {Anandkumar}}]{kovachki2023fno}%
  \BibitemOpen
  \bibfield  {author} {\bibinfo {author} {\bibfnamefont {N.}~\bibnamefont
  {Kovachki}}, \bibinfo {author} {\bibfnamefont {Z.}~\bibnamefont {Li}},
  \bibinfo {author} {\bibfnamefont {B.}~\bibnamefont {Liu}}, \bibinfo {author}
  {\bibfnamefont {K.}~\bibnamefont {Azizzadenesheli}}, \bibinfo {author}
  {\bibfnamefont {K.}~\bibnamefont {Bhattacharya}}, \bibinfo {author}
  {\bibfnamefont {A.}~\bibnamefont {Stuart}},\ and\ \bibinfo {author}
  {\bibfnamefont {A.}~\bibnamefont {Anandkumar}},\ }\href
  {https://doi.org/10.5555/3648699.3648788} {\bibfield  {journal} {\bibinfo
  {journal} {J. Mach. Learn. Res.}\ }\textbf {\bibinfo {volume} {24}},\
  \bibinfo {pages} {1--97} (\bibinfo {year} {2023})}\BibitemShut {NoStop}%
\bibitem [{\citenamefont {Guibas}\ \emph {et~al.}()\citenamefont {Guibas},
  \citenamefont {Mardani}, \citenamefont {Li}, \citenamefont {Tao},
  \citenamefont {Anandkumar},\ and\ \citenamefont
  {Catanzaro}}]{guibas2021afno}%
  \BibitemOpen
  \bibfield  {author} {\bibinfo {author} {\bibfnamefont {J.}~\bibnamefont
  {Guibas}}, \bibinfo {author} {\bibfnamefont {M.}~\bibnamefont {Mardani}},
  \bibinfo {author} {\bibfnamefont {Z.}~\bibnamefont {Li}}, \bibinfo {author}
  {\bibfnamefont {A.}~\bibnamefont {Tao}}, \bibinfo {author} {\bibfnamefont
  {A.}~\bibnamefont {Anandkumar}},\ and\ \bibinfo {author} {\bibfnamefont
  {B.}~\bibnamefont {Catanzaro}},\ }\href@noop {} {\ }\bibinfo {note} {Eprint
  arXiv: 2111.13587 (2021)}\BibitemShut {NoStop}%
\bibitem [{\citenamefont {Lu}\ \emph {et~al.}(2021{\natexlab{b}})\citenamefont
  {Lu}, \citenamefont {Meng}, \citenamefont {Mao},\ and\ \citenamefont
  {Karniadakis}}]{lu2021deepxde}%
  \BibitemOpen
  \bibfield  {author} {\bibinfo {author} {\bibfnamefont {L.}~\bibnamefont
  {Lu}}, \bibinfo {author} {\bibfnamefont {X.}~\bibnamefont {Meng}}, \bibinfo
  {author} {\bibfnamefont {Z.}~\bibnamefont {Mao}},\ and\ \bibinfo {author}
  {\bibfnamefont {G.~E.}\ \bibnamefont {Karniadakis}},\ }\href
  {https://doi.org/10.1137/19m1274067} {\bibfield  {journal} {\bibinfo
  {journal} {SIAM Rev.}\ }\textbf {\bibinfo {volume} {63}},\ \bibinfo {pages}
  {208–228} (\bibinfo {year} {2021}{\natexlab{b}})}\BibitemShut {NoStop}%
\bibitem [{\citenamefont {Pathak}\ \emph {et~al.}()\citenamefont {Pathak},
  \citenamefont {Subramanian}, \citenamefont {Harrington}, \citenamefont
  {Raja}, \citenamefont {Chattopadhyay}, \citenamefont {Mardani}, \citenamefont
  {Kurth}, \citenamefont {Hall}, \citenamefont {Li}, \citenamefont
  {Azizzadenesheli}, \citenamefont {Hassanzadeh}, \citenamefont {Kashinath},\
  and\ \citenamefont {Anandkumar}}]{jaideep2022fcn}%
  \BibitemOpen
  \bibfield  {author} {\bibinfo {author} {\bibfnamefont {J.}~\bibnamefont
  {Pathak}}, \bibinfo {author} {\bibfnamefont {S.}~\bibnamefont {Subramanian}},
  \bibinfo {author} {\bibfnamefont {P.}~\bibnamefont {Harrington}}, \bibinfo
  {author} {\bibfnamefont {S.}~\bibnamefont {Raja}}, \bibinfo {author}
  {\bibfnamefont {A.}~\bibnamefont {Chattopadhyay}}, \bibinfo {author}
  {\bibfnamefont {M.}~\bibnamefont {Mardani}}, \bibinfo {author} {\bibfnamefont
  {T.}~\bibnamefont {Kurth}}, \bibinfo {author} {\bibfnamefont
  {D.}~\bibnamefont {Hall}}, \bibinfo {author} {\bibfnamefont {Z.}~\bibnamefont
  {Li}}, \bibinfo {author} {\bibfnamefont {K.}~\bibnamefont {Azizzadenesheli}},
  \bibinfo {author} {\bibfnamefont {P.}~\bibnamefont {Hassanzadeh}}, \bibinfo
  {author} {\bibfnamefont {K.}~\bibnamefont {Kashinath}},\ and\ \bibinfo
  {author} {\bibfnamefont {A.}~\bibnamefont {Anandkumar}},\ }\href@noop {} {\
  }\bibinfo {note} {Eprint arXiv: 2202.11214 (2022)}\BibitemShut {NoStop}%
\bibitem [{\citenamefont {Jiang}\ \emph {et~al.}()\citenamefont {Jiang},
  \citenamefont {Meinert}, \citenamefont {Jordão}, \citenamefont {Weisser},
  \citenamefont {Holgate}, \citenamefont {Lavin}, \citenamefont {L\"{u}tjens},
  \citenamefont {Newman}, \citenamefont {Wainwright}, \citenamefont {Walker},\
  and\ \citenamefont {Barnard}}]{jiang2021dte}%
  \BibitemOpen
  \bibfield  {author} {\bibinfo {author} {\bibfnamefont {P.}~\bibnamefont
  {Jiang}}, \bibinfo {author} {\bibfnamefont {N.}~\bibnamefont {Meinert}},
  \bibinfo {author} {\bibfnamefont {H.}~\bibnamefont {Jordão}}, \bibinfo
  {author} {\bibfnamefont {C.}~\bibnamefont {Weisser}}, \bibinfo {author}
  {\bibfnamefont {S.}~\bibnamefont {Holgate}}, \bibinfo {author} {\bibfnamefont
  {A.}~\bibnamefont {Lavin}}, \bibinfo {author} {\bibfnamefont
  {B.}~\bibnamefont {L\"{u}tjens}}, \bibinfo {author} {\bibfnamefont
  {D.}~\bibnamefont {Newman}}, \bibinfo {author} {\bibfnamefont
  {H.}~\bibnamefont {Wainwright}}, \bibinfo {author} {\bibfnamefont
  {C.}~\bibnamefont {Walker}},\ and\ \bibinfo {author} {\bibfnamefont
  {P.}~\bibnamefont {Barnard}},\ }\href@noop {} {\ }\bibinfo {note} {Eprint
  arXiv: 2110.07100 (2021)}\BibitemShut {NoStop}%
\bibitem [{\citenamefont {Zhang}, \citenamefont {Benavides-Riveros},\ and\
  \citenamefont {Chen}(2024)}]{zhang2024jpcl}%
  \BibitemOpen
  \bibfield  {author} {\bibinfo {author} {\bibfnamefont {J.}~\bibnamefont
  {Zhang}}, \bibinfo {author} {\bibfnamefont {C.~L.}\ \bibnamefont
  {Benavides-Riveros}},\ and\ \bibinfo {author} {\bibfnamefont
  {L.}~\bibnamefont {Chen}},\ }\href
  {https://doi.org/10.1021/acs.jpclett.4c00598} {\bibfield  {journal} {\bibinfo
   {journal} {J. Phys. Chem. Lett.}\ }\textbf {\bibinfo {volume} {15}},\
  \bibinfo {pages} {3603--3610} (\bibinfo {year} {2024})}\BibitemShut {NoStop}%
\bibitem [{\citenamefont {Ishizaki}\ and\ \citenamefont
  {Tanimura}(2005)}]{ishizaki2005jpsj}%
  \BibitemOpen
  \bibfield  {author} {\bibinfo {author} {\bibfnamefont {A.}~\bibnamefont
  {Ishizaki}}\ and\ \bibinfo {author} {\bibfnamefont {Y.}~\bibnamefont
  {Tanimura}},\ }\href {https://doi.org/10.1143/JPSJ.74.3131} {\bibfield
  {journal} {\bibinfo  {journal} {J. Phys. Soc. Jpn.}\ }\textbf {\bibinfo
  {volume} {74}},\ \bibinfo {pages} {3131--3134} (\bibinfo {year}
  {2005})}\BibitemShut {NoStop}%
\bibitem [{\citenamefont {Hu}, \citenamefont {Xu},\ and\ \citenamefont
  {Yan}(2010)}]{hu2010jcp}%
  \BibitemOpen
  \bibfield  {author} {\bibinfo {author} {\bibfnamefont {J.}~\bibnamefont
  {Hu}}, \bibinfo {author} {\bibfnamefont {R.-X.}\ \bibnamefont {Xu}},\ and\
  \bibinfo {author} {\bibfnamefont {Y.}~\bibnamefont {Yan}},\ }\href
  {https://doi.org/10.1063/1.3484491} {\bibfield  {journal} {\bibinfo
  {journal} {J. Chem. Phys.}\ }\textbf {\bibinfo {volume} {133}},\ \bibinfo
  {pages} {101106} (\bibinfo {year} {2010})}\BibitemShut {NoStop}%
\bibitem [{\citenamefont {Tanimura}(2006)}]{tanimura2006jpsj}%
  \BibitemOpen
  \bibfield  {author} {\bibinfo {author} {\bibfnamefont {Y.}~\bibnamefont
  {Tanimura}},\ }\href {https://doi.org/10.1143/JPSJ.75.082001} {\bibfield
  {journal} {\bibinfo  {journal} {J. Phys. Soc. Jpn.}\ }\textbf {\bibinfo
  {volume} {75}},\ \bibinfo {pages} {082001} (\bibinfo {year}
  {2006})}\BibitemShut {NoStop}%
\bibitem [{\citenamefont {Zhang}\ and\ \citenamefont
  {Tanimura}(2023)}]{zhang2023jcp}%
  \BibitemOpen
  \bibfield  {author} {\bibinfo {author} {\bibfnamefont {J.}~\bibnamefont
  {Zhang}}\ and\ \bibinfo {author} {\bibfnamefont {Y.}~\bibnamefont
  {Tanimura}},\ }\href {https://doi.org/10.1063/5.0156264} {\bibfield
  {journal} {\bibinfo  {journal} {J. Chem. Phys.}\ }\textbf {\bibinfo {volume}
  {159}},\ \bibinfo {pages} {014102} (\bibinfo {year} {2023})}\BibitemShut
  {NoStop}%
\bibitem [{\citenamefont {Kovachki}, \citenamefont {Lanthaler},\ and\
  \citenamefont {Mishra}(2021)}]{kovachki2021fno}%
  \BibitemOpen
  \bibfield  {author} {\bibinfo {author} {\bibfnamefont {N.}~\bibnamefont
  {Kovachki}}, \bibinfo {author} {\bibfnamefont {S.}~\bibnamefont
  {Lanthaler}},\ and\ \bibinfo {author} {\bibfnamefont {S.}~\bibnamefont
  {Mishra}},\ }\href {https://doi.org/10.5555/3546258.3546548} {\bibfield
  {journal} {\bibinfo  {journal} {J. Mach. Learn. Res.}\ }\textbf {\bibinfo
  {volume} {22}},\ \bibinfo {pages} {1--76} (\bibinfo {year}
  {2021})}\BibitemShut {NoStop}%
\bibitem [{\citenamefont {Rosofsky}, \citenamefont {Majed},\ and\ \citenamefont
  {Huerta}(2023)}]{rosofsky2023mlst}%
  \BibitemOpen
  \bibfield  {author} {\bibinfo {author} {\bibfnamefont {S.~G.}\ \bibnamefont
  {Rosofsky}}, \bibinfo {author} {\bibfnamefont {H.~A.}\ \bibnamefont
  {Majed}},\ and\ \bibinfo {author} {\bibfnamefont {E.~A.}\ \bibnamefont
  {Huerta}},\ }\href {https://doi.org/10.1088/2632-2153/acd168} {\bibfield
  {journal} {\bibinfo  {journal} {Mach. learn.: Sci. Technol.}\ }\textbf
  {\bibinfo {volume} {4}},\ \bibinfo {pages} {025022} (\bibinfo {year}
  {2023})}\BibitemShut {NoStop}%
\bibitem [{\citenamefont {Adolphs}\ and\ \citenamefont
  {Renger}(2006)}]{adolphs2006bj}%
  \BibitemOpen
  \bibfield  {author} {\bibinfo {author} {\bibfnamefont {J.}~\bibnamefont
  {Adolphs}}\ and\ \bibinfo {author} {\bibfnamefont {T.}~\bibnamefont
  {Renger}},\ }\href {https://doi.org/10.1529/biophysj.105.079483} {\bibfield
  {journal} {\bibinfo  {journal} {Biophys. J.}\ }\textbf {\bibinfo {volume}
  {91}},\ \bibinfo {pages} {2778--2797} (\bibinfo {year} {2006})}\BibitemShut
  {NoStop}%
\bibitem [{\citenamefont {Ishizaki}\ and\ \citenamefont
  {Fleming}(2009)}]{ishizaki2009pnas}%
  \BibitemOpen
  \bibfield  {author} {\bibinfo {author} {\bibfnamefont {A.}~\bibnamefont
  {Ishizaki}}\ and\ \bibinfo {author} {\bibfnamefont {G.~R.}\ \bibnamefont
  {Fleming}},\ }\href {https://doi.org/10.1073/pnas.0908989106} {\bibfield
  {journal} {\bibinfo  {journal} {Proc. Natl. Acad. Sci.}\ }\textbf {\bibinfo
  {volume} {106}},\ \bibinfo {pages} {17255--17260} (\bibinfo {year}
  {2009})}\BibitemShut {NoStop}%
\bibitem [{\citenamefont {Shi}\ \emph {et~al.}(2009)\citenamefont {Shi},
  \citenamefont {Chen}, \citenamefont {Nan}, \citenamefont {Xu},\ and\
  \citenamefont {Yan}}]{shi2009jcp}%
  \BibitemOpen
  \bibfield  {author} {\bibinfo {author} {\bibfnamefont {Q.}~\bibnamefont
  {Shi}}, \bibinfo {author} {\bibfnamefont {L.}~\bibnamefont {Chen}}, \bibinfo
  {author} {\bibfnamefont {G.}~\bibnamefont {Nan}}, \bibinfo {author}
  {\bibfnamefont {R.-X.}\ \bibnamefont {Xu}},\ and\ \bibinfo {author}
  {\bibfnamefont {Y.}~\bibnamefont {Yan}},\ }\href
  {https://doi.org/10.1063/1.3077918} {\bibfield  {journal} {\bibinfo
  {journal} {J. Chem. Phys.}\ }\textbf {\bibinfo {volume} {130}},\ \bibinfo
  {pages} {084105} (\bibinfo {year} {2009})}\BibitemShut {NoStop}%
\bibitem [{\citenamefont {Yeh}\ and\ \citenamefont {Kais}(2014)}]{yeh2014jcp}%
  \BibitemOpen
  \bibfield  {author} {\bibinfo {author} {\bibfnamefont {S.-H.}\ \bibnamefont
  {Yeh}}\ and\ \bibinfo {author} {\bibfnamefont {S.}~\bibnamefont {Kais}},\
  }\href@noop {} {\bibfield  {journal} {\bibinfo  {journal} {J. Chem. Phys.}\
  }\textbf {\bibinfo {volume} {141}} (\bibinfo {year} {2014})}\BibitemShut
  {NoStop}%
\bibitem [{\citenamefont {Chen}\ \emph {et~al.}(2011)\citenamefont {Chen},
  \citenamefont {Zheng}, \citenamefont {Jing},\ and\ \citenamefont
  {Shi}}]{chen2011jcp}%
  \BibitemOpen
  \bibfield  {author} {\bibinfo {author} {\bibfnamefont {L.}~\bibnamefont
  {Chen}}, \bibinfo {author} {\bibfnamefont {R.}~\bibnamefont {Zheng}},
  \bibinfo {author} {\bibfnamefont {Y.}~\bibnamefont {Jing}},\ and\ \bibinfo
  {author} {\bibfnamefont {Q.}~\bibnamefont {Shi}},\ }\href@noop {} {\bibfield
  {journal} {\bibinfo  {journal} {J. Chem. Phys.}\ }\textbf {\bibinfo {volume}
  {134}} (\bibinfo {year} {2011})}\BibitemShut {NoStop}%
\bibitem [{\citenamefont {Cho}\ \emph {et~al.}(2005)\citenamefont {Cho},
  \citenamefont {Vaswani}, \citenamefont {Brixner}, \citenamefont {Stenger},\
  and\ \citenamefont {Fleming}}]{cho2005jpcb}%
  \BibitemOpen
  \bibfield  {author} {\bibinfo {author} {\bibfnamefont {M.}~\bibnamefont
  {Cho}}, \bibinfo {author} {\bibfnamefont {H.~M.}\ \bibnamefont {Vaswani}},
  \bibinfo {author} {\bibfnamefont {T.}~\bibnamefont {Brixner}}, \bibinfo
  {author} {\bibfnamefont {J.}~\bibnamefont {Stenger}},\ and\ \bibinfo {author}
  {\bibfnamefont {G.~R.}\ \bibnamefont {Fleming}},\ }\href
  {https://doi.org/10.1021/jp050788d} {\bibfield  {journal} {\bibinfo
  {journal} {J. Phys. Chem. B}\ }\textbf {\bibinfo {volume} {109}},\ \bibinfo
  {pages} {10542–10556} (\bibinfo {year} {2005})}\BibitemShut {NoStop}%
\bibitem [{\citenamefont {Borrelli}(2019)}]{borrelli2019jcp}%
  \BibitemOpen
  \bibfield  {author} {\bibinfo {author} {\bibfnamefont {R.}~\bibnamefont
  {Borrelli}},\ }\href {https://doi.org/10.1063/1.5099416} {\bibfield
  {journal} {\bibinfo  {journal} {J. Chem. Phys.}\ }\textbf {\bibinfo {volume}
  {150}},\ \bibinfo {pages} {234102} (\bibinfo {year} {2019})}\BibitemShut
  {NoStop}%
\bibitem [{\citenamefont {Borrelli}\ and\ \citenamefont
  {Gelin}(2021)}]{borrelli2021wirescms}%
  \BibitemOpen
  \bibfield  {author} {\bibinfo {author} {\bibfnamefont {R.}~\bibnamefont
  {Borrelli}}\ and\ \bibinfo {author} {\bibfnamefont {M.~F.}\ \bibnamefont
  {Gelin}},\ }\href {https://doi.org/10.1002/wcms.1539} {\bibfield  {journal}
  {\bibinfo  {journal} {WIREs Comput. Mol. Sci.}\ }\textbf {\bibinfo {volume}
  {11}},\ \bibinfo {pages} {e1539} (\bibinfo {year} {2021})}\BibitemShut
  {NoStop}%
\bibitem [{\citenamefont {Wang}\ and\ \citenamefont
  {Thoss}(2004)}]{wang2004chemphyslett}%
  \BibitemOpen
  \bibfield  {author} {\bibinfo {author} {\bibfnamefont {H.}~\bibnamefont
  {Wang}}\ and\ \bibinfo {author} {\bibfnamefont {M.}~\bibnamefont {Thoss}},\
  }\href {https://doi.org/10.1016/j.cplett.2004.03.052} {\bibfield  {journal}
  {\bibinfo  {journal} {Chem. Phys. Lett.}\ }\textbf {\bibinfo {volume}
  {389}},\ \bibinfo {pages} {43--50} (\bibinfo {year} {2004})}\BibitemShut
  {NoStop}%
\bibitem [{\citenamefont {Wang}\ and\ \citenamefont
  {Thoss}(2008)}]{wang2008chemphys}%
  \BibitemOpen
  \bibfield  {author} {\bibinfo {author} {\bibfnamefont {H.}~\bibnamefont
  {Wang}}\ and\ \bibinfo {author} {\bibfnamefont {M.}~\bibnamefont {Thoss}},\
  }\href {https://doi.org/10.1016/j.chemphys.2007.12.004} {\bibfield  {journal}
  {\bibinfo  {journal} {Chem. Phys.}\ }\textbf {\bibinfo {volume} {347}},\
  \bibinfo {pages} {139--151} (\bibinfo {year} {2008})}\BibitemShut {NoStop}%
\end{thebibliography}%

\end{document}